\documentclass{aa}  

\usepackage{graphicx}
\usepackage{txfonts}

\def\co2{CO$_2$}
\def\ch4{CH$_4$}
\def\h2{H$_{2}$}
\def\h2o{H$_2$O}
\def\hdu18{HD~189733~b}
\def\hd20{HD~209458~b}
\def\hep{He$^{+}$}
\def\hes{He(1$^{1}$S)}
\def\het{He(2$^{3}$S)}
\def\mlr{$\dot M$}
\def\lya{Ly$\alpha$}

\def\T{{\cal T}}
\def\A{{\cal A}}
\def\ve{{\rm v}}
\def\rd{{\rm d}}

\def\kms{km\,s$^{-1}$}
\def\cms{cm$^{-3}$\,s$^{-1}$}

\def\rp{$R_{\rm P}$}
\def\rj{$R_{\rm Jup}$\ }
\def\mj{$M_{\rm Jup}$\ }
\def\vwind{v$_{wind}$}

\def\gs{g\,s$^{-1}$}
\newcommand{\overbar}[1]{\mkern 1.5mu\overline{\mkern-1.5mu#1\mkern-1.5mu}\mkern 1.5mu}

\begin{document} 

\title{Modelling the He~{\sc i} triplet absorption at 10830\,{\AA} in the atmosphere of HD\,209458\,b}

%
%
\titlerunning{Modelling of He(2$^{3}$S) in HD~209458~b}
\author{M.~Lamp{\'o}n\inst{1}, 
M.~L\'opez-Puertas\inst{1}, L.~M.~Lara\inst{1}, A.~S\'anchez-L\'opez\inst{1}, M.~Salz\inst{2}, S.~Czesla\inst{2}, 
J.~Sanz-Forcada\inst{3}, K.~Molaverdikhani\inst{4}, F.~J.~Alonso-Floriano\inst{5}, L.~Nortmann\inst{6,7}, J.~A.~Caballero\inst{3}, F.~F.~ Bauer\inst{1},  E.~Pall{\'e}\inst{6,7}, D.~Montes\inst{8}, A.~Quirrenbach\inst{9}, E.~Nagel\inst{2}, I.~Ribas\inst{10,11}, A.~Reiners\inst{12}, P.~J.~Amado\inst{1}. 
}

\institute{Instituto de Astrof{\'i}sica de Andaluc{\'i}a (IAA-CSIC), Glorieta de la Astronom{\'i}a s/n, 18008 Granada, Spain\\
\email{mlampon@iaa.es}
\and
Hamburger Sternwarte, Universit{\"a}t Hamburg, Gojenbergsweg 112, 21029 Hamburg, Germany
\and
Centro de Astrobiolog{\'i}a (CSIC-INTA), ESAC, Camino bajo del castillo s/n, 28692 Villanueva de la Ca{\~n}ada, Madrid, Spain
\and
Max-Planck-Institut f{\"u}r Astronomie, K{\"o}nigstuhl 17, 69117 Heidelberg, Germany
\and
Leiden Observatory, Leiden University, Postbus 9513, 2300 RA, Leiden, The Netherlands
\and
Instituto de Astrof{\'i}sica de Canarias, Calle V{\'i}a L{\'a}ctea s/n, 38200 La Laguna, Tenerife, Spain
\and
Departamento de Astrof{\'i}sica, Universidad de La Laguna, 38026  La Laguna, Tenerife, Spain
\and
Departamento de F{\'i}sica de la Tierra y Astrof{\'i}sica \& IPARCOS-UCM (Instituto de F{\'i}sica de Part{\'i}culas y del Cosmos de la UCM), Facultad de Ciencias F{\'i}sicas, Universidad Complutense de Madrid,  28040 Madrid, Spain
\and
Landessternwarte, Zentrum f\"ur Astronomie der Universit\"at Heidelberg, K\"onigstuhl 12, 69117 Heidelberg, Germany
\and
Institut de Ci\`encies de l'Espai (CSIC-IEEC), Campus UAB, c/ de Can Magrans s/n, 08193 Bellaterra, Barcelona, Spain
\and
Institut d'Estudis Espacials de Catalunya (IEEC), 08034 Barcelona, Spain
\and
Institut f{\"u}r Astrophysik, Georg-August-Universit{\"a}t, Friedrich-Hund-Platz 1, 37077 G{\"o}ttingen, Germany
}

\authorrunning{Lamp\'on et al.}
\date{Received 22 November 2019 / Accepted 21 February 2019}

\abstract
{\hd20 is an exoplanet with an upper atmosphere undergoing blow-off escape that has  mainly  been studied using measurements of the \lya\ absorption.
Recently, high-resolution measurements of absorption in the He~{\sc i} triplet line at 10830\,{\AA} of several exoplanets (including \hd20) have been reported, creating a new opportunity to probe escaping atmospheres. 
}
{We aim to better understand the atmospheric regions of \hd20 from where the escape originates.}
{We developed a 1D hydrodynamic model with spherical symmetry for the \hd20 thermosphere coupled with a non-local thermodynamic model for the population of the He~{\sc i} triplet state. 
In addition, we performed high-resolution radiative transfer calculations of synthetic spectra for the He triplet lines and compared them with the measured absorption spectrum in order to retrieve information about the atmospheric parameters.}
{We find that the measured spectrum constrains the [H]/[H$^{+}$] transition altitude occurring in the range of 1.2\,\rp\ to 1.9\,\rp. Hydrogen is almost fully ionised at altitudes above 2.9\,\rp. We also find that the X-ray and extreme ultraviolet absorption takes place at effective radii from 1.16 to 1.30\,\rp, and that the He~{\sc i} triplet peak density occurs at altitudes from 1.04 to 1.60\,\rp. Additionally, the averaged mean molecular weight is confined to the  0.61--0.73\,g\,mole$^{-1}$ interval, and the thermospheric H/He ratio should be larger than 90/10, and most likely approximately 98/2. We also provide a one-to-one relationship between mass-loss rate and temperature. 
Based on the energy-limited escape approach and assuming heating efficiencies of 0.1--0.2, we find a mass-loss rate in the range of (0.42--1.00)\,$\times$\,$10^{11}$\,\gs\ and a corresponding temperature range of 7125\,K to 8125\,K.}
{The analysis of the measured He~{\sc i} triplet absorption spectrum significantly constrains the thermospheric structure of \hd20 and advances our knowledge of its escaping atmosphere.}

\maketitle
%

\section{Introduction}
\label{Intro}
 
Close-in exoplanets receive considerable high-energy irradiation that triggers an outgoing bulk motion and expands their atmospheres. This in turn results in a very efficient mechanism for atmospheric mass loss. In some cases, the velocity reaches supersonic values that give rise to the so-called blow-off escape. This escape process is key for understanding the formation and evolution of observed (exo)planets. 
\cite{Lecavelier_des_Etangs_2004} showed that such an efficient escape could lead to the loss of a significant fraction of the atmosphere in hot Jupiters, or even to a complete atmospheric loss in some hot Neptunes. In the same vein, \cite{Locci_2019} showed that the occurrence of gas giants of less than 2\,\mj in young stars is higher than in old ones. 
Furthermore, evaporation escape has also been invoked to explain the distribution with respect to the star--planet separation of super-Earths and mini-Neptunes \citep{Jin_2018}. 
Observations of extended atmospheres and their modelling are therefore essential for understanding the physical conditions that trigger evaporation escape and for obtaining a deeper understanding of planetary diversity, formation, and evolution.        
 
The first detection of an evaporating atmosphere was reported by \cite{VidalMadjar2003}, who observed the \lya\ line in absorption in the atmosphere of \hd20 and concluded that the signal originated from escaping hydrogen. 
Subsequent studies, both theoretical \citep[e.g.,][]{Yelle_2004,Tian2005,Penz_2008,san10,san11,Koskinen2013a,Salz2016}
and observational  
\citep{VidalMadjar2003,VidalMadjar2004,Vidal_Madjar_2008,Vidal_Madjar_2013,Ballester_2007,Ballester_2015,Linsky2010,Ehrenreich_2008,Ben_Jaffel_2010,Jensen_2012}, 
supported the prediction that the extended atmosphere of \hd20\ can be explained by a highly effective escape. Following this thread, hydrodynamic escape models were used to explain the observations of the absorption excess in \lya\  and ultraviolet (UV) in other planets, such as in HD~189733~b, WASP-12\,b, GJ~436~b, GJ~3470~b, and KELT-9~b \citep{Vidal_Madjar_2013,Ehrenreich_2008,Ehrenreich_2015,Lecavelier_des_Etangs_2010, Lecavelier_des_Etangs_2012,Fossati_2010, Haswell_2012, Bourrier_2013, Ben_Jaffel_2013, Kulow_2014, Lavie_2017,Yan_2018}.     

\lya\ observations are restricted to space-based telescopes and are significantly affected by absorption in the interstellar medium, which leads to a scarcity of observational data. The He\,{\sc i} $ 2^3S-2^3P$ triplet, hereafter \het, composed of three lines at 10830.33, 10830.25, and 10829.09\,{\AA}, is not strongly absorbed by the interstellar medium \citep{Indriolo_2009} and can be observed from the ground. \cite{Seager_2000} studied transmission spectra for close-in planets and estimated significant absorption of \het\ that could be observable, particularly when the planet has an extended atmosphere. More recently, \cite{Oklopcic2018} developed a model suitable for estimating \het\ absorption in extended atmospheres. 

\het\ absorption was detected for the first time in the atmosphere of WASP-107 b with the Wide Field Camera 3 (WFC3) onboard the {\em Hubble Space Telescope} \citep{Spake_2018}, and almost simultaneously  with the high-resolution spectrograph CARMENES \cite[Calar Alto high-Resolution search for M dwarfs with Exoearths with Near-infrared and optical Échelle Spectrographs;][]{Quirrenbach16, Quirrenbach18} at the 3.5\,m Calar Alto Telescope in the atmospheres of WASP--69~b \citep{Nortmann2018}, HAT--P--11~b \citep{Allart2018}, and HD~189733~b \citep{Salz2018}. A few more observations were also recently performed. Thus, HAT--P--11~b data were acquired with WFC3 \citep{Mansfield_2018}, whereas CARMENES was also used to observe WASP-107~b \citep{Allart_2019} and \hd20 \citep{Alonso2019}. 

In this work we analyse the new \het\ observations of \hd20 reported by \cite{Alonso2019}. Based on the fact that this absorption probes the region where the escape originates, we aim at gaining a better understanding of the atmosphere of \hd20 and  establishing tighter constraints on its mass-loss rate, temperature, and composition (e.g. the H/He ratio). To this end, we developed a 1D hydrodynamic and spherically symmetric model 
together with a non-local thermodynamic model for the \het\ state similar to that reported by \cite{Oklopcic2018}. The overall model computes the \het\ radial distribution that it is then entered into a high-resolution line-by-line radiative transfer model in order to compute \het\  absorption profiles. Synthetic spectra are then compared with the observed one in order to derive the properties of the escaping atmosphere.

This paper is organised as follows: Sect.\,\ref{observ} summarises the observations and Sect.\,\ref{atm_model} describes the modelling of the \het\ radial densities and its absorption. The results, and a discussion on comparisons with previous works for the mass-loss rate, temperature, and H density, are presented in Sect.~\ref{results}. The main conclusions are summarised in Sect.~\ref{conclusions}.

\section{Observations of He~{\sc i} $\lambda$10830\,{\AA}\ absorption}
\label{observ}

\begin{figure}
\includegraphics[angle=90, width=1.0\columnwidth]{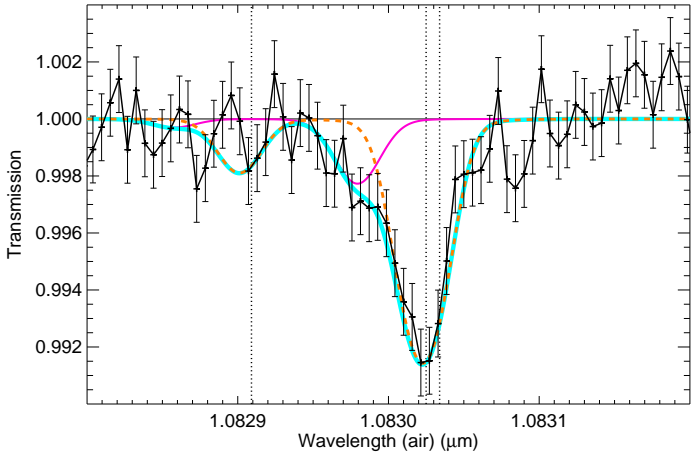}\hspace*{0.35cm}
\caption{Spectral transmission of the He triplet at mid transit. Data points and their respective error bars are shown in black (adapted from \cite{Alonso2019}). The orange dash curve is the best fit obtained for a temperature of 6000\,K, a mass-loss rate ($\dot{M}$) of 1.9\,$\times\,10^{9}$\,\gs\ , and an H/He mole-fraction ratio of 90/10 (see Sect.~\ref{results}). We note that these values have changed with respect to those reported in \citeauthor{Alonso2019} The magenta line is a modelled absorption of the helium triplet moving along the observer's line of sight with $-$13\,\kms\ with respect to the rest-frame of the planet and using the appropriate \het\ density to fit the measured spectrum. The cyan curve is the total modelled absorption. The positions of the three helium lines are marked by vertical dotted lines.}  
\label{absorption}
\end{figure}

Here we briefly summarise the observations reported by \cite{Alonso2019} of the helium excess absorption from the atmosphere of \hd20 using CARMENES (see Fig.~\ref{absorption}).
The He excess absorption peaks at a value of 0.91$\pm$0.10\% at mid-transit.
The core of the absorption shows a net blueshift of 1.8$\pm$1.3\,\kms, suggesting that the helium envelope moves at that velocity, in the line of sight from the Earth with respect to the rest-frame of the planet.
This velocity is rather low, similar to that reported by \citet{Snellen2010} from observations of the carbon monoxide band near 2\,$\mu$m that occurs in the lower thermosphere, and suggests that the He triplet absorption  takes place at relatively low altitudes (within a few planetary radii, see Sect.~\ref{results} below). 

Figure\,\ref{absorption} also shows excess absorption near 10829.8\,\AA. Since the absorption is well above the estimated errors, it might be caused by an additional atmospheric absorption component. However, it is largely blue shifted, by around 13\,\kms, and therefore could be due to atmospheric escape occurring beyond the thermosphere. 
\cite{Alonso2019} constructed the light curve of the He~{\sc i} signal (see their Fig.~5) 
and obtained an average in-transit absorption of $\sim$0.44\%, about a factor two smaller than the peak  absorption. However, there is no clear evidence for pre- or post-transit absorption signals.

We assumed a one-dimensional hydrodynamic and spherically symmetric model for analysing this absorption. 
{This simplification is justified because}: 1) the bulk of the absorption takes place at relatively low radii, well below the Roche lobe height, and with rather low wind velocities; 2) there are no clear signs of asymmetry in the absorption; and 3) the light curve does not exhibit signs of pre- or post-transit absorption -- for example there are no cometary-like tail features. 
This model  also has the advantage that it is computationally very efficient   
with respect to 1D models that solve the energy balance equation, and is  about one order of magnitude faster than the model of \cite{Garcia_munoz_2007} (priv. comm.) and about two orders of magnitude faster than the model of \cite{Salz2016}. It also allows us to explore a wide range of atmospheric parameters, and yet gives reasonable density distributions (see, e.g. the comparison with the model of \citet{Salz2016} in Sect.\,\ref{model_val} and Fig.\,\ref{den_cand} below). 

The analysis of the velocity shifts in the \het\ absorption is potentially interesting because it would provide us information about the 3D velocity distribution and maybe also break some of the degeneracy between temperature and mass-loss rates. Our model, as it is one dimensional and spherically symmetric, cannot explain net blue or red shifts and therefore such an analysis is beyond the scope of this paper.  Nevertheless, in order to better fit the measured spectrum we assume a \ velocity of $-$1.8\,\kms along the observational line of sight superimposed on the radial velocities of our model. In the following sections we describe the He triplet density model and the radiative absorption calculations used to constrain the characteristics of the escaping atmosphere of this planet.

\section{Modelling of the helium triplet density and the absorption} \label{atm_model}

\begin{table}
\centering
\caption{\label{table.parameters}System parameters of HD\,209458.}
\begin{tabular}{l c l} 
\hline  \hline  \noalign{\smallskip}
Parameter & Value & Reference  \\
\noalign{\smallskip} \hline \noalign{\smallskip}
$R_{\star}$ &   1.155\,$^{+0.014}_{-0.016}$\,$R_\sun$    & \citet{Torres08}     \\
\noalign{\smallskip}
$M_{\star}$ & 1.119\,$\pm$\,0.033\,$M_\sun$     & \citet{Torres08}      \\
\noalign{\smallskip}
$T_{\rm eff}$ & 6065\,$\pm$\,50\,K      & \citet{Torres08}      \\
\noalign{\smallskip}
$[\rm Fe/H]_{\star}$ & 0.02\,$\pm$\,0.05        & \citet{Santos_2004}   \\
\noalign{\smallskip}
$a$             & 0.04707\,$^{+0.00046}_{-0.00047}$\,au & \citet{Torres08} \\
\noalign{\smallskip}
$R_{P}$         & 1.359\,$^{+0.016}_{-0.019}$\,\rj\     & \citet{Torres08}      \\
\noalign{\smallskip}
$M_{P}$         & 0.685\,$^{+0.015}_{-0.014}$\,\mj\     & \citet{Torres08}      \\
\noalign{\smallskip}
\hline
\end{tabular}
\end{table}

We computed the \het\ radial density by means of a 1D\ hydrodynamic and spherically symmetric model \citep[see, e.g.][]{Yelle_2004,Tian2005,Garcia_munoz_2007,Koskinen2013a,Salz2016} and following the methods discussed in \cite{Oklopcic2018}. The hydrodynamic equations for mass and momentum conservation, for a steady-state radial atmospheric outflow are given by 
\begin{equation}
        \frac{d\ (r^2\, \rho(r)\, \ve(r)) }{d r} = 0, 
        \label{eq:mass_cons}
\end{equation}
and
\begin{equation}
        \ve(r)\,  \frac{d\ve}{dr} + \frac{1}{\rho(r)} \frac{dp}{dr} + \frac{G\, M_P}{r^2} = 0, 
        \label{eq:momentum_cons}
\end{equation}
where $r$ is the distance from the centre of the planet; $\rho$ and $\ve$ are the mass density and bulk radial velocity of the gas, respectively; $p$ is the gas pressure; $G$ the gravitational constant; and $M_P$ is the planet mass (see Table\,\ref{table.parameters}). The momentum equation contains only the forces induced by the gravity of the planet and the gas pressure, neglecting other forces such as the stellar gravitational pull and fluid viscosity. 
We also consider the atmosphere as an ideal gas, $p = \rho\, k\,T/\mu$, where $k$ is Boltzmann's constant, $T$ is temperature,  and $\mu$ is the gas mean molecular weight. 
The pressure gradient can then be written as
\begin{equation}
    \frac{dp}{dr} = \frac{k\,T(r)}{\mu (r)} \frac{d\rho}{dr} + \,\rho(r)\,\frac{d}{dr} \left ( \frac{k\,T(r)}{\mu(r)} \right ).
    \label{eq:deriv_press}
\end{equation}

The derivative of Eq.~(\ref{eq:mass_cons}) gives 
\begin{equation}
     \frac{1}{\rho(r)}\, \frac{d\rho }{dr} = - \frac{1}{\ve(r)}\,\frac{d\ve}{dr}-\frac{2}{r},
    \label{eq:interm}
\end{equation}
and including Eqs.~(\ref{eq:deriv_press}) and (\ref{eq:interm}) into the momentum continuity equation (Eq.~\ref{eq:momentum_cons}), this becomes 
\begin{equation}
    \begin{split}
     \ve(r) \frac{d\ve}{dr}\,
     &+\,\ve^{2}_s(r)\,\left( - \frac{1}{\ve(r)}\,\frac{d\ve}{dr}-\frac{2}{r}  \right) \\
     &+\,\frac{d(\ve^{2}_s(r))}{dr}
     + \frac{G\,M_{p}}{r^{2}} = 0,
     \end{split}
    \label{eq:momentum_cons_2}
\end{equation}
where $\ve_s(r) = \left[ k\,T(r)/\mu (r) \right]^{1/2}$ is the gas speed of sound.

To solve Eq.~(\ref{eq:momentum_cons_2}) we need to incorporate the energy and species continuity equations into the system, 
which requires the use of computationally expensive numerical methods. Our aim is to develop a very fast hydrodynamic model in order to explore and constrain the mass-loss rate and temperatures imposed by the \het\, absorption measurements. 
To that end, we assume here a constant speed of sound, $\ve_{\rm s,0}$, which allows us to decouple the momentum equation from the energy budget equation and obtain a simple expression for  Eq.~(\ref{eq:momentum_cons_2}) that can be solved analytically. This expression is known as the isothermal Parker wind approximation, previously studied by \cite{Parker1958} for describing stellar winds, and can be written as
\begin{equation}
    \begin{split}
     \frac{1}{\ve(r)} \frac{d\ve}{dr}\,
     \left( \frac{\ve^{2}(r)}{\ve^{2}_{\rm s,0}}\, - 1 \right) 
     = \frac{2}{r}- \frac{G\,M_{p}}{\ve^{2}_{\rm s,0}\,r^{2}}.
     \end{split}
    \label{eq:momentum_cons_3}
\end{equation}
However, our approach does not require temperature to be necessarily constant, but to have the same inverse altitude dependence as $\mu(r)$ so that the ratio $T(r)/\mu(r)$ is constant, i.e. $\ve_{s,0} = \left[ k\,T(r)/\mu (r) \right]^{1/2}$. 

We obtain the constant speed of sound from
\begin{equation}
     \ve_{\rm s,0} =  \left( \frac{k\,T_0}{\overbar\mu} \right)^{1/2}, \label{eq:constant_vs}
\end{equation}
where we introduce a constant temperature $T_0$ and the corresponding average mean molecular weight $\overbar\mu$ of the gas. This quantity is calculated using  Eq.\,(\ref{eq:mmw_averaged}), which is constructed such that it provides an accurate calculation of the integrals over velocity and radius of the hydrodynamic equations.

Below we discuss (see Sect.~\ref{model_val}) the validity and scope of this approach. In particular,  the total density of our model closely matches that obtained by more comprehensive models when the constant speed of sound is equal to the maximum speed of sound of those models. Under these conditions, the $T_0$ of our model is very close to the maximum of the temperature profile obtained by those models.

Equation (\ref{eq:momentum_cons_3}) has infinite solutions but only one has a physical interpretation in terms of escape: the transonic solution that describes a subsonic velocity of the gas at distances below the sonic point (the altitude where wind velocity is equal to the speed of sound) and supersonic beyond that point \citep[see, e.g.][]{Parker1958,Lamers1999}. Integrating Eq.~(\ref{eq:momentum_cons_3}) and selecting the solutions that cross the sonic point, we have 
 \begin{equation}
         \frac{\ve(r)}{\ve_{\rm s,0}} \exp \left[ - \frac{\ve^2(r)}{2\ve_{\rm s,0}^2} \right] = \left( \frac{r_s}{r} \right)^{2} \exp \left(- \frac{2r_s}{r} +   \frac{3}{2} \right),
         \label{eq:velocity_profile}
   \end{equation}
where $r_s$\,=\,$ G\, M_P/2\,\ve_{\rm s,0}^2$ is the radial distance of the sonic point. Of the two possible solutions of Eq.~(\ref{eq:velocity_profile}) the transonic one is the solution that provides the velocity profile of the hydrodynamic escape.  

Integrating Eq.~(\ref{eq:mass_cons}) and taking into account spherical symmetry, the mass-loss rate, $\dot{M}$, can be expressed by
\begin{equation}
         \dot M = 4\,\pi\, r^{2}\, \rho(r)\, \ve(r),
         \label{eq:mass_loss_rate}
   \end{equation}
and including the velocity profile from Eq.~(\ref{eq:velocity_profile}), the density profile results in
\begin{equation}
         \frac{\rho(r)}{\rho_s} = \text{exp} \left[ \frac{2r_s}{r} - \frac{3}{2} - \frac{\ve^2(r)}{2\ve_{\rm s,0}^2} \right],
         \label{eq:density_profile}
   \end{equation}
where $\rho_s$ is the gas density at the sonic point. 

\begin{table*}
\centering
\caption{Production and loss processes included in the \het\ model. \label{table.proc}}
\begin{tabular}{l l l l} 
\hline \hline\noalign{\smallskip}
Name & Process & Rate$^{a}$ & Reference  \\
  \noalign{\smallskip}
\hline \noalign{\smallskip}
$J_H$   & H + h$\nu$            $\rightarrow$ H$^+$ + e$^-$     & see text & \cite{Osterbrock_2006}        \\
$\alpha_H$      & H$^+$ + e$^-$                 $\rightarrow$ H                     & 2.59 $\times 10^{-13} (T/10^{4})^{-0.700}$         & \cite{Osterbrock_2006}            \\
$J_{\rm He^{1}S}$       & He(1$^1$S) + h$\nu$           $\rightarrow$ He$^+$ + e$^-$         & see text      & \cite{Brown_1971}    \\
$\alpha_1$      & He$^+$ + e$^-$        $\rightarrow$ He(1$^1$S)                & 1.54 $\times 10^{-13}  (T/10^{4})^{-0.486}$     & \cite{Benjamin_1999}     \\
$J_{\rm He^{3}S}$       & He(2$^3$S) + h$\nu$ $\rightarrow$ He$^+$ + e$^-$         & see text      & \cite{Norcross_1971}      \\
$\alpha_3$      & He$^+$ + e$^-$        $\rightarrow$ He(2$^3$S)                & 2.10 $\times 10^{-13} (T/10^{4})^{-0.778}$      & \cite{Benjamin_1999}  \\
$q_{13}$        & He(1$^1$S) + e$^-$            $\rightarrow$ He(2$^3$S) + e$^-$& 2.10 $\times 10^{-8} \, \sqrt{\frac{13.6}{kT}} \, $exp$ \left(-\frac{19.81}{kT} \right)\Upsilon_{13}$   & \cite{Bray_2000,Oklopcic2018}     \\
$q_{31a}$       & He(2$^3$S) + e$^-$  $\rightarrow$ He(2$^1$S) + e$^-$& 2.10 $\times 10^{-8} \, \sqrt{\frac{13.6}{kT}} \, $exp$ \left(-\frac{0.80}{kT} \right)\frac{\Upsilon_{31a}}{3}$        & \cite{Bray_2000,Oklopcic2018} \\
$q_{31b}$       & He(2$^3$S) + e$^-$  $\rightarrow$ He(2$^1$P) + e$^-$& 2.10 $\times 10^{-8} \, \sqrt{\frac{13.6}{kT}} \, $exp$ \left(-\frac{1.40}{kT} \right)\frac{\Upsilon_{31b}}{3}$        & \cite{Bray_2000,Oklopcic2018}\\
$Q_{31}$        & He(2$^3$S) + H        $\rightarrow$ He(1$^1$S) + H             &       5.00 $\times 10^{-10}$  & \cite{Roberge_1982,Oklopcic2018} \\
$Q_{{\rm He}^+}$        & He$^+$ + H    $\rightarrow$ He(1$^1$S) + H$^+$                 &       1.25 $\times 10^{-15}(\frac{300}{T})^{-0.25}$   & \cite{Glover_2007,Koskinen2013a} \\
$Q_{\rm He}$    & He(1$^1$S) + H$^+$    $\rightarrow$ He$^+$ + H                 &       1.75 $\times 10^{-11}(\frac{300}{T})^{0.75}$ exp$(\frac{-128000}{T})$         & \cite{Glover_2007,Koskinen2013a} \\
$A_{31}$        & He(2$^3$S)            $\rightarrow$ He(1$^1$S) + h$\nu$                 &       1.272 $\times 10^{-4}$  &  \cite{Drake_1971,Oklopcic2018}     \\
\noalign{\smallskip}\hline
\end{tabular}
\tablefoot{\tablefoottext{a}{Units are [\cms] for the recombination and collisional processes and [s$^{-1}$] for the other processes. $\Upsilon_{ij}$ are the effective collision strengths taken from \citep{Bray_2000} at the corresponding temperatures.}}
\end{table*}

The radial distribution of the species H, H$^+$, helium singlet, hereafter \hes, \hep, and \het, are obtained by solving their respective continuity equations,
\begin{align}
         &-\ve(r)\ \frac{ \partial f_i}{ \partial r} + P_i + L_i = 0,
         \label{eq:continuity}
   \end{align}
where the first term accounts for advection, $f_i$ is the fraction of species $i$ (i.e. with respect to the total ---neutral, ionised, and excited atom concentrations--- of species $i$), $P_i$ is the production, and $L_i$ is the loss term.

The production and loss terms and corresponding rates included for the different species are essentially those considered by \cite{Oklopcic2018}, but two additional processes have been included: the charge exchange reactions, $Q_{\rm He}$ and $Q_{{\rm He}^+}$, which were taken from \cite{Koskinen2013a}; see Table\,\ref{table.proc}.
 
Thus, production and loss incorporate photo-ionisation and recombination processes of atomic hydrogen (processes $J_H$ and $\alpha_H$ in Table\,\ref{table.proc}). For helium, these processes include photo-ionisation ($J_{\rm He^{1}S}$ and $J_{\rm He^{3}S}$), recombination ($\alpha_1$ and $\alpha_3$), collisional excitation and de-excitation  with electrons ($q_{13}$ and $q_{31}$), collisional de-activation with H atoms ($Q_{31}$), charge exchange between atoms and ions ($Q_{\rm He}$ and $Q_{{\rm He}^+}$), and the radiative relaxation of the helium triplet state ($A_{31}$). 
Double ionisation processes for helium were not considered, and we assume that the charge exchange reactions between helium and hydrogen do not significantly affect the concentration of the latter. Furthermore, we assume that the electrons produced from helium ionisation are negligible in comparison to the production from hydrogen ionisation.

\begin{figure}
\includegraphics[angle=90.0,width=1\columnwidth]{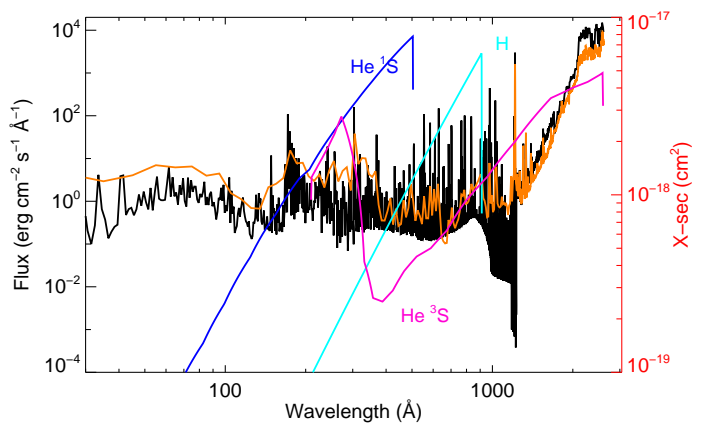}\hspace*{0.35cm}
\caption{ Flux density for HD\,209458 at 0.04707\,au (left y-axis) as calculated here (black) and for the modern Sun (solar maximum) as reported by \citet{Claire2012} and scaled up to the size of HD\,209458 (orange). We note that the resolution of the two spectra is different. The H, He singlet, and He triplet ionisation cross sections (right y-axis) are also shown.} 
\label{flux_xsec} 
\end{figure}

The hydrogen photo-ionisation cross sections were taken from \cite{Osterbrock_2006} and the photo-ionisation cross sections for the helium triplet and singlet states were taken from \cite{Norcross_1971} and \cite{Brown_1971}, respectively; (see Fig.~\ref{flux_xsec}). 
The photo-ionisation rates are calculated as usual,
\begin{align}
         &J(r) = \int_{\lambda_{min}}^{\lambda_{max}} \sigma_{\lambda}\, F_{\lambda}\, \exp[- \tau_{\lambda}(r)]\ d\lambda, 
         \label{photo-ionisation}
\end{align}
and 
\begin{align}
         & \tau_{\lambda}(r) = \sum_i \int_{r}^{\infty} \sigma_{i}(\lambda) \, n_i(r) \, \rd r,
         \label{tau}
\end{align}
where $\lambda_{min}$ and $\lambda_{max}$ correspond to the wavelength range of photo-ionisation, $\tau_{\lambda}(r)$ is the optical depth, $F_{\lambda}$ is the stellar flux density at the top of the atmosphere, and $\sigma_i(\lambda)$ and $n_i(r)$ are the absorption cross section and number density
of species $i$ (H and He), respectively.    
  
The flux of HD\,209458 in the 5--1230\,\AA\ range was calculated using a coronal model based on {\em XMM-Newton} observations of this star \citep{san10,san11} that are combined to improve their statistical significance in the spectral fit, reaching a 3.2--$\sigma$ detection. No substantial variability was registered in the X-rays observations, although the number of measurements is small. 
The coronal 1-T model, $\log T$(K)$=6.0^{+0.3}_{-0.0}$, $\log EM$\,(cm$^{-3}$)\,=\,49.52$^{+0.22}_{-0.48}$, $L_{\rm X}$\,=\,6.9$\times10^{26}$\,erg\,s$^{-1}$, is complemented with line fluxes from observations with the Cosmic Origin Spectrograph onboard the {\em Hubble Space Telescope} ({\em HST}) 
published by \citet{fran10} in order to extend the model towards lower temperatures ($\log T$(K)$\sim$ 4.0--6.5). 
No substantial amount of emitting material is expected for higher temperatures.
The XUV (5--920\,\AA) modelled luminosity is 1.5$\times10^{28}$\,erg\,s$^{-1}$, which is about 2.6 times higher than in \cite{san11}.
For the 1230--1700\,\AA\ range we used current observations with the Space Telescope Imaging Spectrograph taken from the {\em HST} archive\footnote{STIS/G140L combined spectra were acquired on the dates 9 Oct 2003, 19 Oct 2003, 6 Nov 2003 and 24 Nov 2003.}. At $\lambda > 1700$\,\AA\ we used the stellar atmospheric model of \citet{CK2004} scaled to the temperature, surface gravity, and metallicity of HD\,209458 (see Table\,\ref{table.parameters}). The resulting stellar flux for the spectral range 5--2600\,\AA\ at the orbital separation of the planet (0.04707\,au) is shown in Fig.~\ref{flux_xsec}.  
This flux shows several differences with respect to the solar flux \citep[see, e.g.][]{Claire2012,Linsky2014}. Thus, it shows a weaker \lya\ continuum  but more prominent emission lines in the 600-912\,\AA\ region. Also, the continuum in the  950--1200\,\AA\ range is significantly weaker, although this is expected to have a minor effect on the \het\ population. In order to analyse the effects of these phenomena on the \het\ density we performed some tests as shown in Sect.\,\ref{density}.

Equations \ref{eq:velocity_profile}, \ref{eq:density_profile}, \ref{eq:mmw_averaged}, and \ref{eq:continuity} are solved iteratively until convergence is reached, ensuring a fully consistent model. As the abundances of neutral and ionised species and electrons vary with $r$, so does the mean molecular weight. We assume pre-set neutral gas composition of H and He (we considered H/He mole-fraction ratios of 90/10, 95/5, and 98/2). To ease the numerical convergence of the model,  we select the minimum value of $\overbar{\mu}$ (atmosphere fully ionised in the first iteration).
The $\rho(r)$ and $\ve(r)$ profiles derived from the solution of Eqs.~\ref{eq:velocity_profile} and \ref{eq:density_profile} with this initial $\overbar{\mu}$ value enter into the system of continuity equations  (Eqs.~\ref{eq:continuity}) to obtain the abundances of the neutral and ionised species, from which we calculated a new mean molecular weight profile, $\mu(r)$.
The sequence is repeated until the $\mu(r)$ profiles in two subsequent iterations differ by less than 1\%. 
In our model we assumed the lower boundary to be the pressure level of 1 mbar, corresponding roughly to a distance of 1.04\,\rp\ or $\sim$3800\,km above \rp=1 (\rp=1 at 1 bar).

\subsection{Validity of the hydrodynamic model}\label{model_val}

\begin{figure}
\includegraphics[angle=0.0, width=1.0\columnwidth]{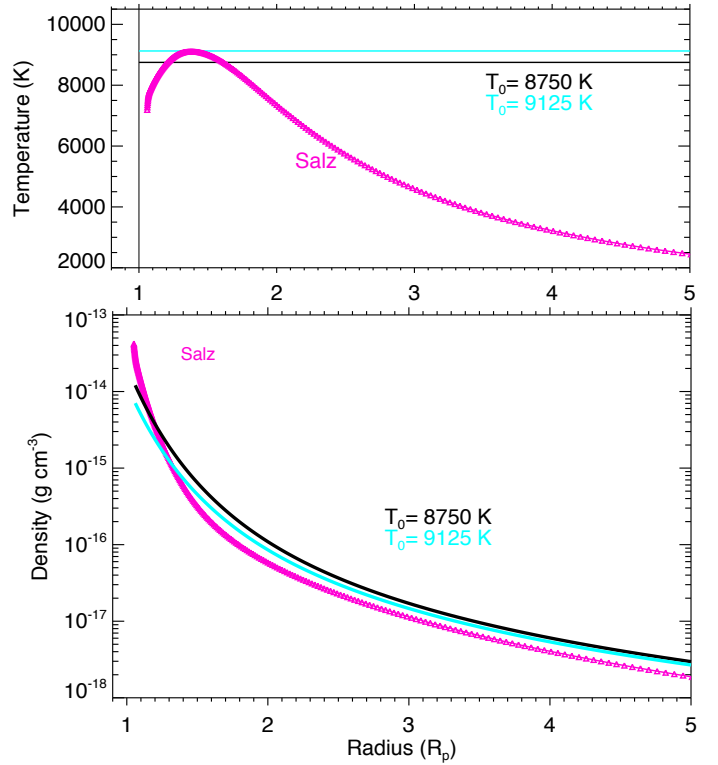}
\caption{Comparison of temperatures (upper panel) and their corresponding density profiles (lower panel). {\em Upper panel}: Black line is the temperature resulting from the maximum speed of sound of \cite{Salz2016} and our averaged mean molecular weight; the cyan line is the maximum of the temperature profile of \citeauthor{Salz2016}, and magenta is their temperature profile. 
{\em Lower panel}: Density profiles calculated by \citeauthor{Salz2016} (magenta) and those calculated using our model (black and cyan). 
} \label{den_cand} 
\end{figure}

In order to check the validity of our constant-speed-of-sound approach, we compared our results with those obtained by \cite{Salz2016}, who solved the energy equation and did not include any assumption on temperature.  For that comparison, we used the same \hd20 bulk parameters, mass-loss rate, H/He, and XUV stellar flux. We performed the calculations for two $T_0$ temperatures: (1) that obtained when considering the largest speed of sound computed by \cite{Salz2016} and the average mean molecular weight obtained in our model (Eq.\,\ref{eq:constant_vs}); and (2) the maximum of the temperature profile computed by \cite{Salz2016}. Figure\,\ref{den_cand} shows these latter two temperature profiles and the one calculated by \cite{Salz2016}, as well as the corresponding total density profiles. Except for altitudes very close to the lower boundary (r$<$1.15\,\rp), the density that we obtain with the temperature corresponding to the maximum speed of sound in \cite{Salz2016}, 8750\,K, agrees with that computed by these latter authors. This result should not be surprising because for the maximum of the speed of sound, its derivative is zero and then the equations of the full model reduce to those in our model, i.e. Eq.~\ref{eq:momentum_cons_2} reduces to Eq.~\ref{eq:momentum_cons_3}.

The temperature $T_0$\,=\,8750\,K is also close to the thermospheric maximum temperature calculated by \cite{Salz2016}. Moreover, the density obtained for the maximum of the temperature profile of \citeauthor{Salz2016}, $T_0$\,=\,9125\,K,  still gives very good agreement with that of \citeauthor{Salz2016}, although at altitudes close to the lower boundary, where the \het\ absorption is most important (see Fig.\,\ref{he3_candidates}), the agreement is slightly poorer. Furthermore, the temperature profile obtained in our approach, dictated by $\ve_{s, \rm max}(\rm{Salz}) = \left[ k\,T(r)/\mu (r) \right]^{1/2}$, is equal to $T_0$\,=\,8750\,K at an altitude close to 1.6\,\rp, which corresponds to the location of the photo-ionisation front, that is, where most of the stellar radiation is absorbed.
We also performed this comparison with two other planets, HD 189733b and GJ 3470b, which show different parameters and maximum of their temperature profiles at significantly different altitudes \citep{Salz2016}. Since we obtain similar results, the input temperature to our model, $T_0$, can be considered as a good proxy (within $\sim$10\%) for the maximum temperature obtained in the hydrodynamic models that  also solve the energy budget equation.

\begin{figure}
\includegraphics[angle=90, width=1.0\columnwidth]{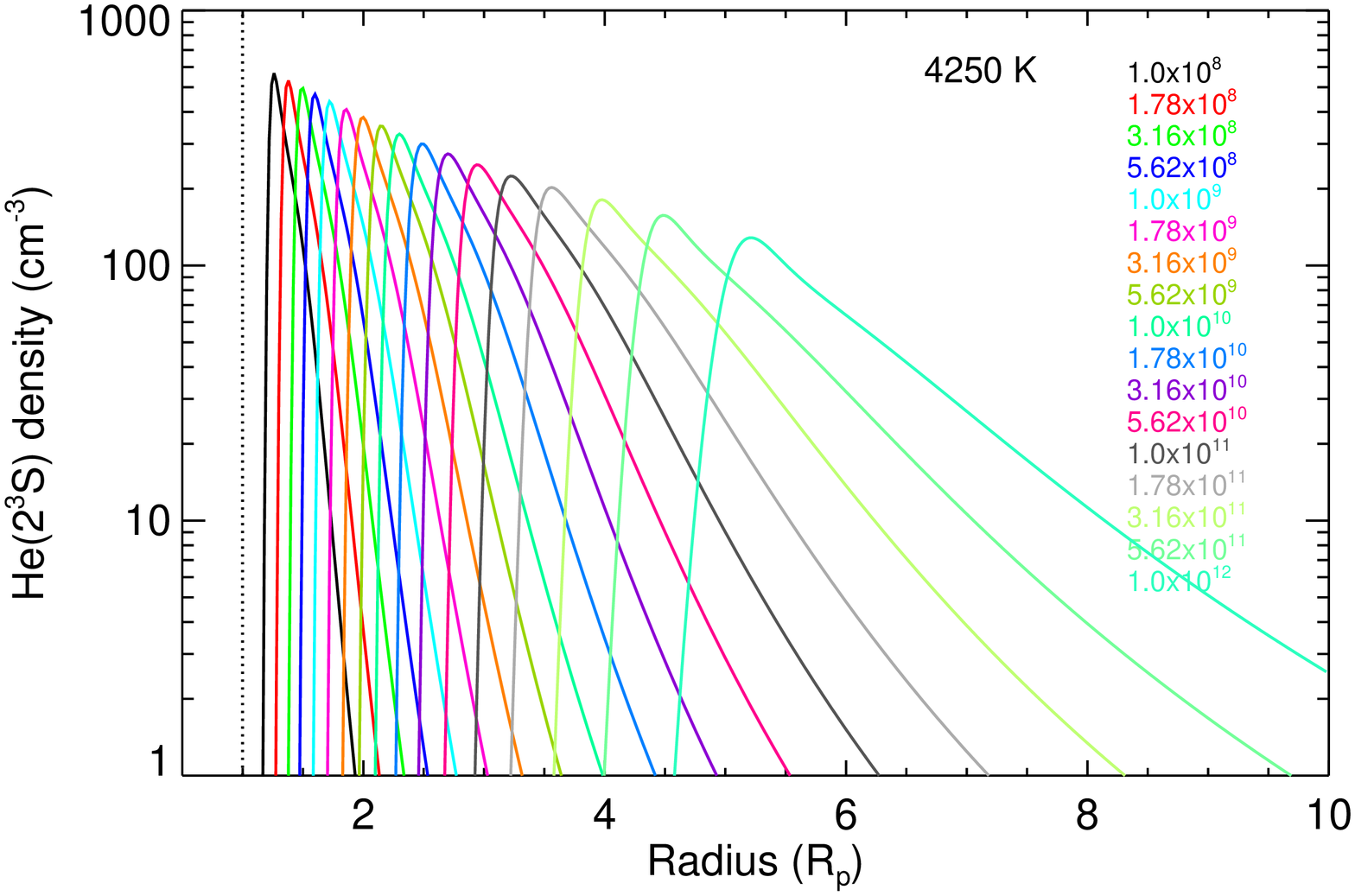}

 \includegraphics[angle=90, width=1.0\columnwidth]{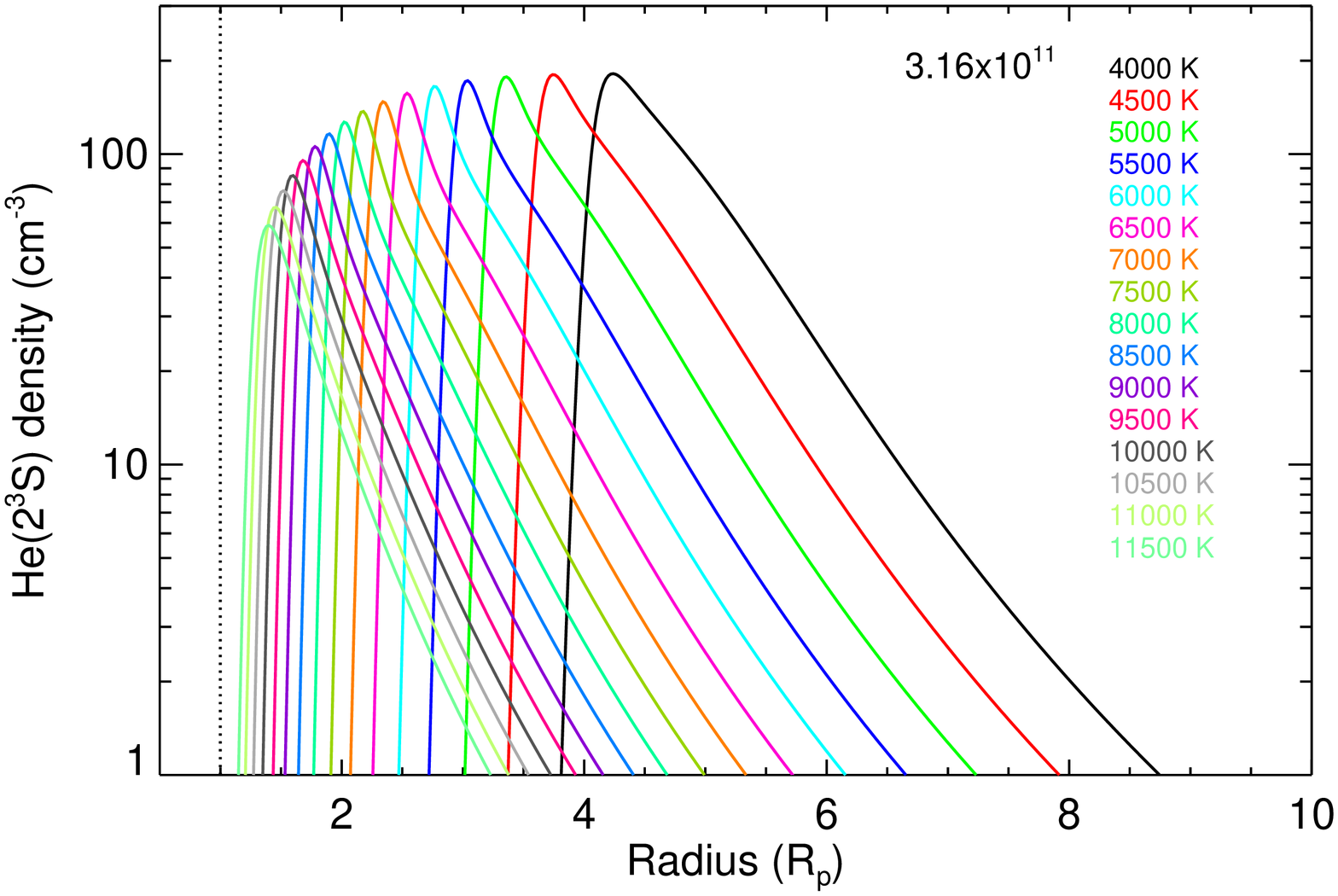} 
\caption{Density profiles of the helium triplet state showing the mass-loss rate dependence for a temperature of 4250\,K (top panel) and the temperature dependence for a mass-loss rate of 3.16$\times$\,10$^{11}$\gs (bottom panel). An H/He ratio of 90/10 was assumed.} \label{fig_He3density} 
\end{figure}

\subsection{\het\ densities}\label{density}

Figure\,\ref{fig_He3density} shows several \het\ density profiles and their dependence on the mass-loss rate (upper panel) and temperature (lower panel) for an H/He ratio of 90/10. As a general trend, for a given temperature, smaller mass-loss rates give rise to higher
\het\ peak densities and more compressed (narrower  layers of) \het\ density profiles. Conversely, at larger mass-loss rates, the \het\ density has a broader shape and its maximum occurs at larger planetary radii. Regarding the effect of temperature, for a given $\dot M$, we can see that the peak of the \het\ density occurs at smaller radii for larger temperatures, whereas colder thermospheres produce higher \het\ abundances. 

\begin{figure}
\includegraphics[angle=90.0, width=1.0\columnwidth]{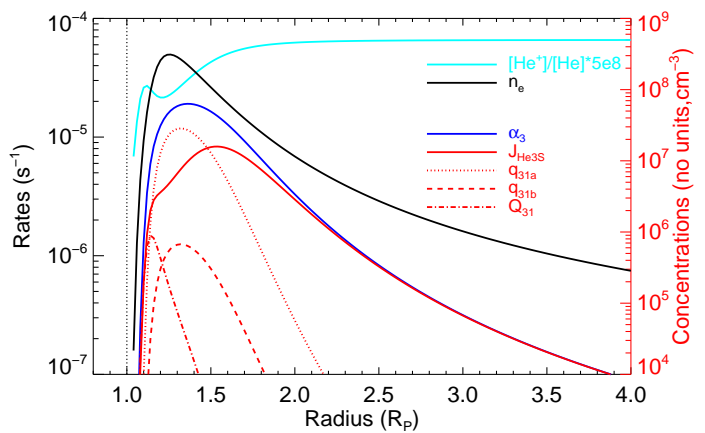} 
\caption{Production and loss rates of \het\ (labelled as in Table\,\ref{table.proc}, left y-axis) for the case of 6000\,K and a mass-loss rate of 4.2$\times\,10^{9}$\,\gs. The blue line corresponds to the production by recombination and the red lines to the losses. Also shown are the concentrations of the species directly involved in the recombination (right-y axis): the electrons (in cm$^{-3}$, black solid line) and the [\hep]/[He] ratio (scaled by 5$\times\,10^{8}$ in cyan). An H/He ratio of 90/10 was assumed.} 
\label{termsHe3} 
\end{figure}

To understand this behaviour we analysed the production and loss terms of Eq.~(\ref{eq:continuity}). As an example, Fig.~\ref{termsHe3} shows the production and loss terms of \het\ for a temperature of 6000\,K and a mass-loss rate of 4.2$\times\,10^{9}$\,\gs.
The production of \het\ is dominated by the recombination of \hep\ with electrons,  $\alpha_3$, having its maximum contribution at about 1.2--1.4\,\rp. 
The electron density is mainly driven by H photo-ionisation, $J_H$, whereas the electron productions from the photo-ionisation of \hes\ and \het\ are negligible. In turn, the \hep\ production is determined by the helium photo-ionisation, $J_{\rm He^1S}$ , and the charge exchange process,  $Q_{\rm He}$; and its losses are mainly controlled by the recombination with electrons, $\alpha_1$ and $\alpha_3$, and the charge exchange process $Q_{{\rm He}^+}$. At high altitudes, where He ionisation is very effective, the \hep\ concentration is dominated by its photo-ionisation, $J_{\rm He^1S}$, but at lower altitudes the photo-ionisation production and the losses by recombination with electrons  determine the resulting \hep\ profile (see Fig.~\ref{termsHe3}). 
The production of \het\ by collisional excitation, $q_{13}$, is very small, i.e. below the lower limit of the scale of Fig.~\ref{termsHe3}. We found that the inclusion of the charge-exchange processes $Q_{\rm He}$ and $Q_{{\rm He}^+}$ produces a net loss of \hep, which translates into a reduction of the \het\ concentration peak ($\sim$15\% for this case).

\begin{figure}
\includegraphics[angle=90, width=1.0\columnwidth]{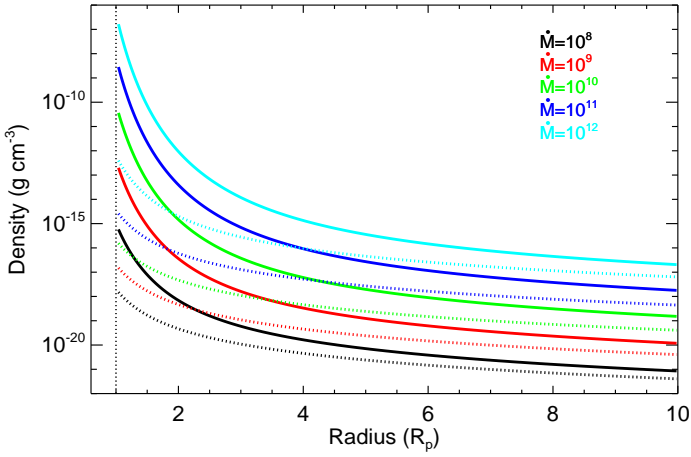}
\caption{Total atmospheric density profiles for several mass-loss rates and temperatures of 5000\,K (solid) and 10000\,K (dotted). An H/He ratio of 90/10 was assumed.} \label{fig:density} 
\end{figure}

Given the importance of H photo-ionisation in the production of \het, it can be seen that a thinner atmosphere (i.e. weaker $\dot M$ or warmer thermosphere, see Fig.\ref{fig:density}) produces ionisation of H at lower altitudes and hence effectively produces \het\ at this region. In contrast, for larger $\dot M$ or colder thermospheres, the density is higher and the absorption of the stellar flux takes place mainly at higher altitudes. Additionally we observe in Fig.\,\ref{fig_He3density} a broader extension of  [\het] (square brackets indicate concentration) at larger $\dot M$, which is due to the weaker vertical gradient of the total density at higher altitudes (see Fig.\ref{fig:density}). For these conditions, the attenuation of the stellar flux occurs more progressively with altitude than at low altitudes where the density abruptly increases (see Fig.~\ref{fig:density}). In addition to this, the hotter the gas is, the more expanded and thinner the thermosphere is.   

Losses of \het\ are controlled by different processes. The major loss is due to photo-ionisation, $J_{He^3S}$, at higher altitudes and the collisional deactivation with electrons, $q_{31a}$, at lower altitudes. Collisional deactivation with H, $Q_{31}$, and advection are about one order of magnitude smaller than photo-ionisation and collisional deactivation with electrons. Overall, the helium triplet concentration is mainly controlled by the stellar flux: on the one hand, by the electron production from H photo-ionisation, which heavily depends on the stellar flux at 600--912\,\AA\ (see Fig.~\ref{flux_xsec}), and on the other hand, by the losses through photo-ionisation of \het, which mainly depends on the stellar flux at $\sim$1000-2600\,\AA\ (see Fig.~\ref{flux_xsec}). Hence, the ratio of the stellar flux in these two regions, 600-912\,\AA\ and $\sim1000-2600$ \AA, will impact the \het\ density favouring a larger \het\ density as the ratio increases. 

As discussed above, we performed some tests  using the solar flux scaled to the size of HD\,209458 (see Fig.\,\ref{flux_xsec}) for temperatures ranging from 6000 to 10000\,K and \mlr  \ from 10$^{9}$ to 10$^{11}$\,\gs. When using this flux, the production terms of \het, i.e. the photo-ionisation rates $J_H$ and $J_{\rm He^{1}S}$, change by factors of $\sim$0.46 and $\sim$2.3, respectively, while the loss term, $J_{\rm He^{3}S}$, decreases by a factor of $\sim$0.44; overall, this produces an increase in the \het\ concentration of a factor of approximately 2.4.

Different \het\ density profiles are obtained when considering different H/He ratios, mass-loss rates, or temperatures. However, the same mechanism described above applies. The helium triplet abundance is largely determined by the stellar flux via H photo-ionisation as well as by the density, which largely controls the atmospheric region where the stellar flux is fully absorbed.

\subsection{Spectral absorption}
\label{transmission}

\begin{figure}
\centering
\includegraphics[angle=90,width=\hsize]{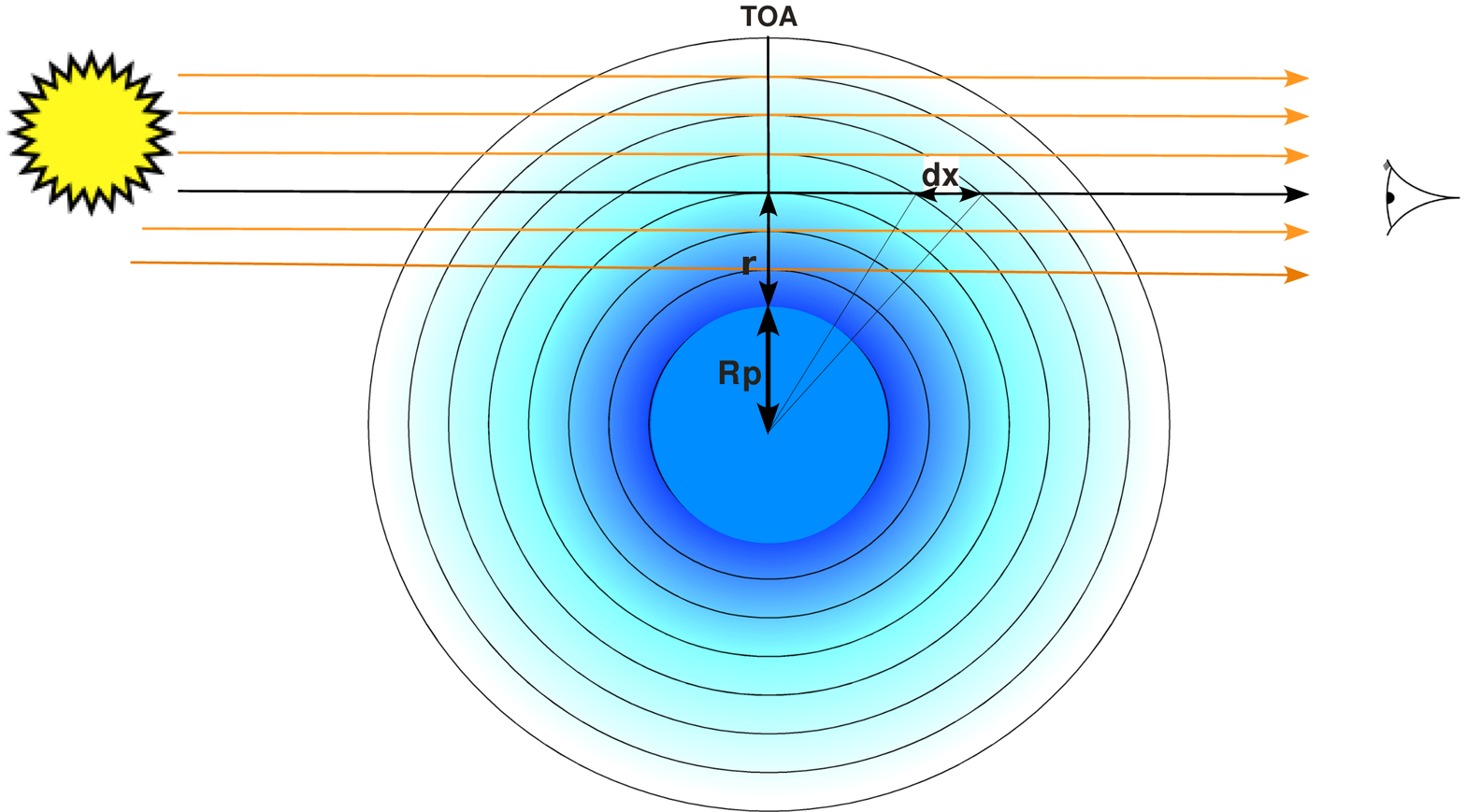}
\includegraphics[angle=90, width=\hsize]{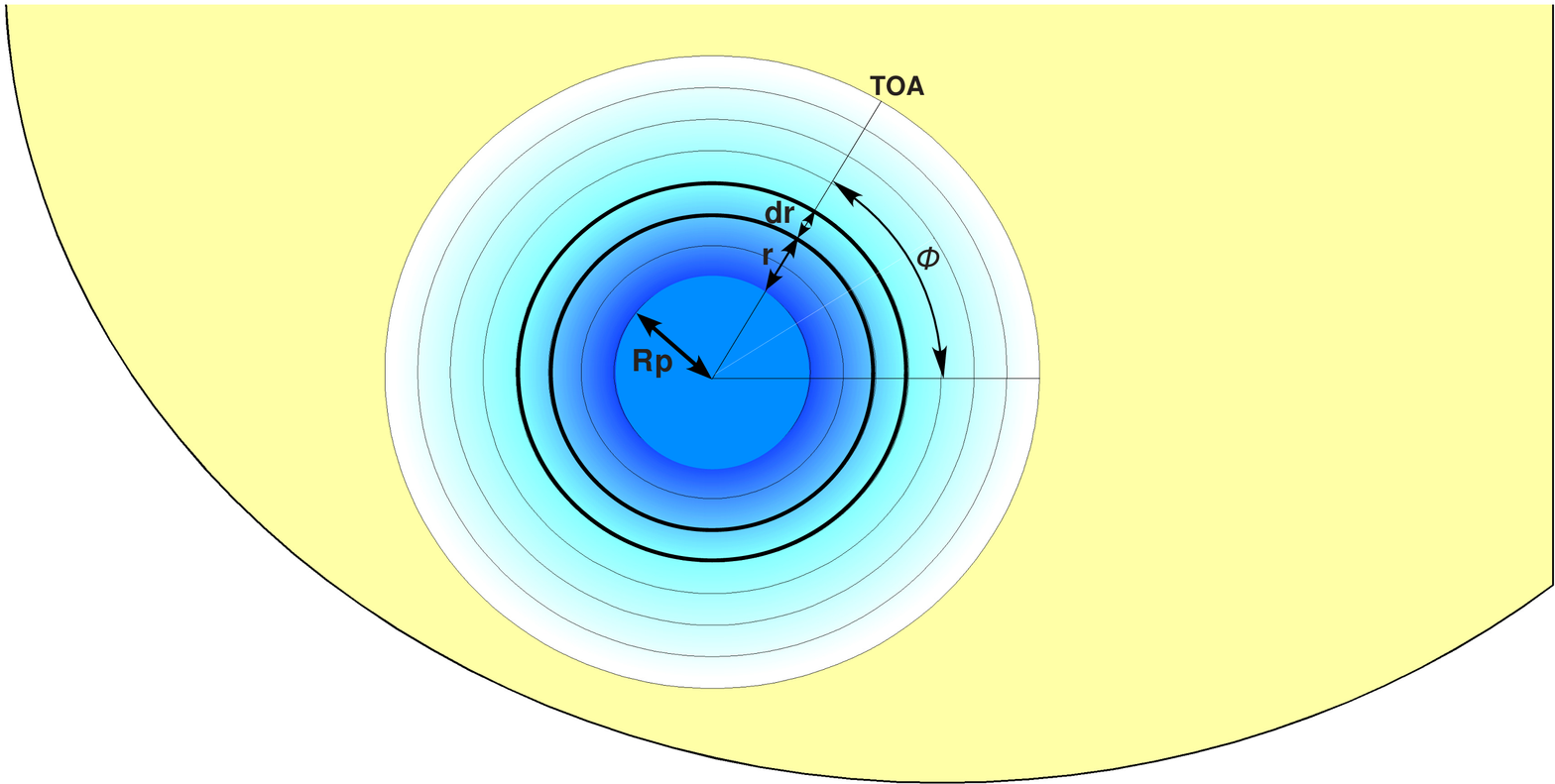}
\caption{Basic sketch (not to scale) of the side view (top) and front view (bottom) of the typical geometry of the primary transit.}
\label{sketch}
\end{figure}

The spectral absorption of the He triplet lines was calculated computing the radiative transfer following the usual transit geometry (see Fig.~\ref{sketch}). The transmission, $\T_{\nu}(r)$, along the line of sight (LOS) $x$ at a radius $r$ over the planet reference surface for an infinitesimal field of view of the planet's atmosphere (see Fig.~\ref{sketch}, top) at frequency $\nu$ can be written as \citep[see, e.g.,][p. 64]{Lopez-Puertas2001}
\begin{equation}
  \T_{\nu}(r) = \exp\left[-\int_{-TOA}^{TOA} k_{\nu}(x)\ n(x)\ \rd x\ \right],  
\end{equation}
where $k_{\nu}$ is the absorption coefficient of the radiative transition and $n(x)$ is the concentration of the absorbing gas, the helium triplet state in our case, and $TOA$ stands for the top of the atmosphere.

The absorption coefficient can be expressed as $k_{\nu}=K\ f_{\nu}$, where $K$ is the integrated absorption coefficient and $f_{\nu}$ is the line shape. The latter is usually taken as a Lorentz, Doppler, or Voigt profile. Here, as most of the absorption comes from low-pressure regions, we assumed a Doppler profile which in general accounts for temperature and the turbulent broadening, for example
\begin{equation}
f_{\nu}(x) = \frac{1}{\alpha_D\,\sqrt{\pi}} \exp\Bigg[ - \frac{[\nu-\nu_0+(\nu_0/c)\,{\rm v}_{wind}]^2}{\alpha_D^2}\Bigg], \label{eq:doppler}
\end{equation}
where $c$ is the speed of light and $\nu_0$ the central frequency of the line. The Doppler line width, $\alpha_D$, is given by
\begin{equation}
\alpha_D= \frac{\nu_0}{c}\sqrt{\frac{2kT}{m}+ {\rm v}^2_{turb}}~~~{\rm with}~~~{\rm v}^2_{turb}=\frac{5kT}{3m}, 
\end{equation}
where ${\rm v}^2_{turb}$ is the turbulent velocity, and $m$ is the mass of the atom or molecule. We note that in Eq.~(\ref{eq:doppler}) we have included
a term with \vwind, the mean velocity of the gas along the line of sight (towards the observer), in order to account for possible motion of the absorbing gas. This dependence can take place either along the LOS ray pencil, $x$ (top panel of Fig.~\ref{sketch}), or along the radial and azimuth dependencies of the \vwind, $r$ and $\phi$ (lower panel of Fig.~\ref{sketch}).

The absorption of the whole atmosphere, $\A_{\nu}$, is obtained by integration of the absorption for the area covered by infinitesimal spherical rings of thickness $\rd r$ and then integrating over all radii $r$ and polar angles $\phi$ of the atmosphere,
\begin{equation}
  \A_{\nu} = \int_{0}^{2\pi} \int_{0}^{TOA} (R_p+r)\ [1-\T_{\nu}(r,\phi)]\  \rd r\  \rd \phi,   \label{eq:abs}
\end{equation}
where $R_p$ is the radius of the planet.
We include the integration over the polar angle, $\phi$, in order to account for potential atmospheric inhomogeneities in the density or velocity of the helium triplet state along this coordinate. 

The absorption coefficients, $K$, and the frequencies, $\nu_0$, for the three lines from meta-stable helium levels were taken from the NIST Atomic Spectra Database (https://www.nist.gov/pml/atomic-spectra-database). The \het\ density profiles were calculated as described above in this section. 
Temperature was kept constant, as given by the helium triplet density model. As most of the absorption is expected to take place at atmospheric regions below the Roche limit (see below), here we did not include the broadening of the lines due to turbulence (${\rm v}_{turb}$\,=\,0 in Eq.~\ref{eq:doppler}).
The \vwind\ included in the nominal calculation corresponds to the line-of-sight component of the radial velocity outflow obtained in the density model of Sect.~\ref{atm_model}. 

\section{Results and discussion} \label{results}

To analyse the mid-transit absorption spectra of Fig.~\ref{absorption}, the thermospheric model described in Sect.~\ref{atm_model} has been run for a temperature range of 4000\,K to 11500\,K in steps of 125\,K. The mass-loss rate interval is 10$^{8}$\,g\,s$^{-1}$ to 10$^{12}$\,g\,s$^{-1}$ in eight steps per decade. This results in a total of 2013 simulations for each set of H/He abundances. 

\begin{figure}
\includegraphics[angle=90, width=1.0\columnwidth]{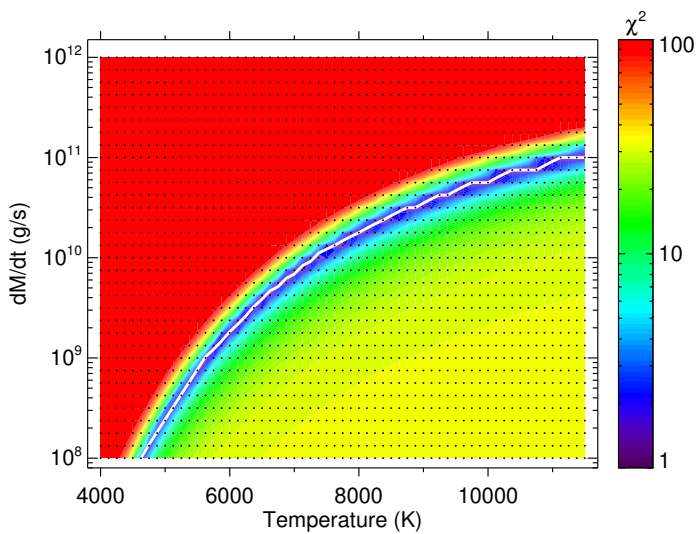} 
\caption{Contour map of the $\chi^2$ of the model of the helium triplet absorption compared to that measured by CARMENES as reported by \citet{Alonso2019}. The $\chi^2$ has been scaled by 10$^5$ (or, equivalently, the reduced $\chi^2$ scaled by a factor of 3.8). An H/He ratio of 90/10 was assumed. The
small black dots represent the grid of models. The white curve highlights the best fitting simulations.}  
\label{chi2}
\end{figure}

\begin{figure}
\includegraphics[angle=90.0, width=1.0\columnwidth]{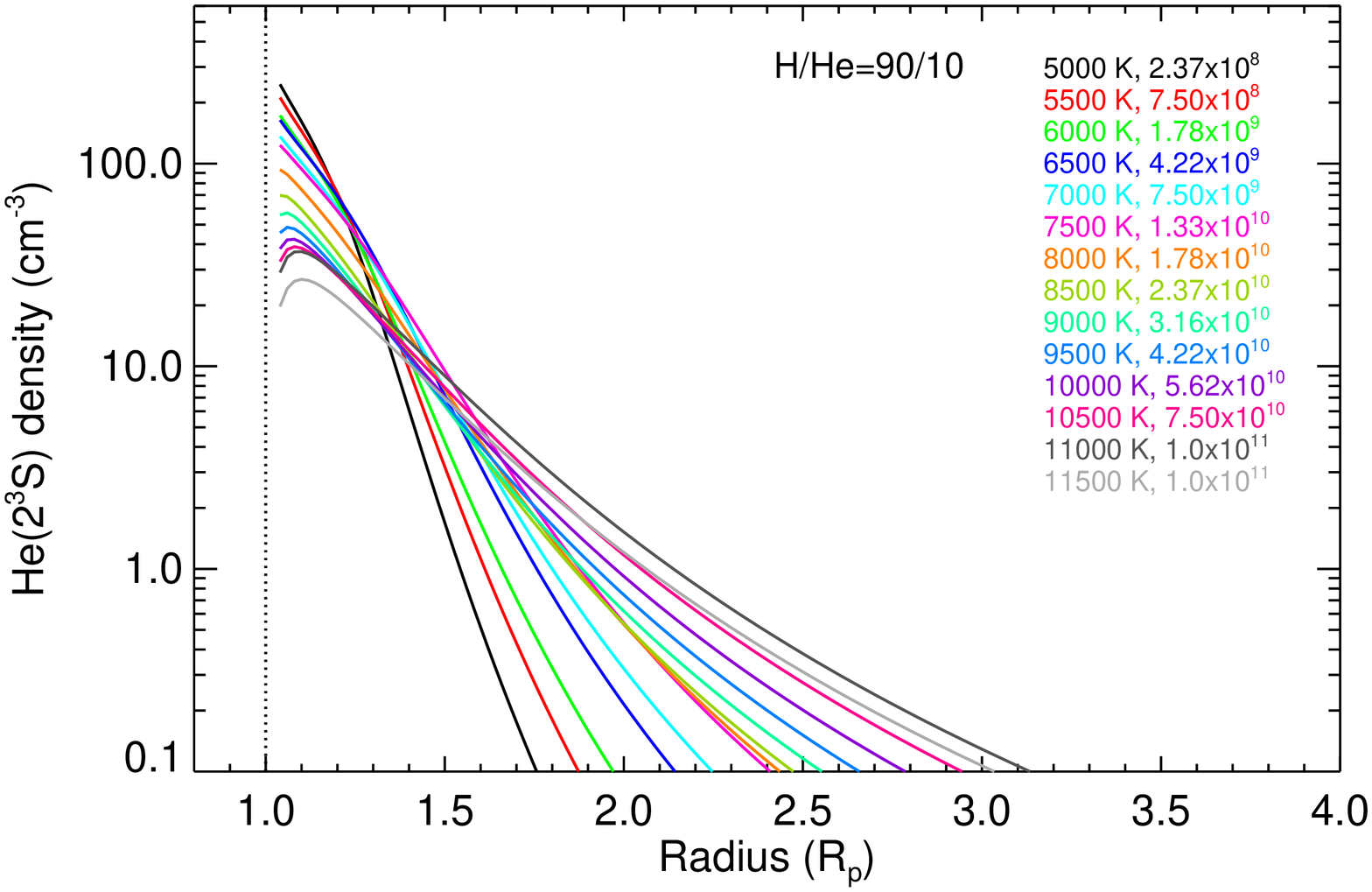}
\includegraphics[angle=90.0, width=1.0\columnwidth]{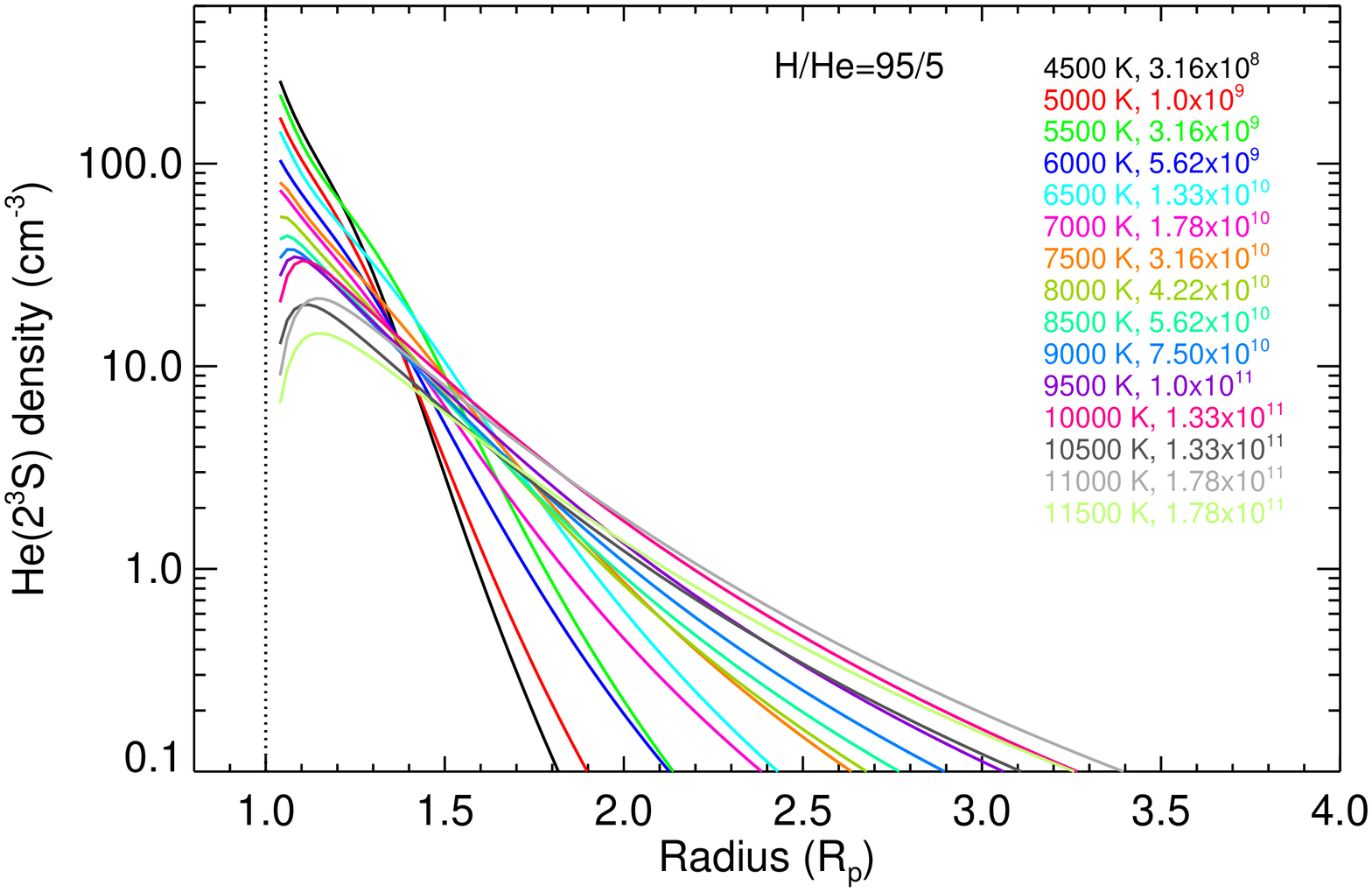}
\includegraphics[angle=90.0, width=1.0\columnwidth]{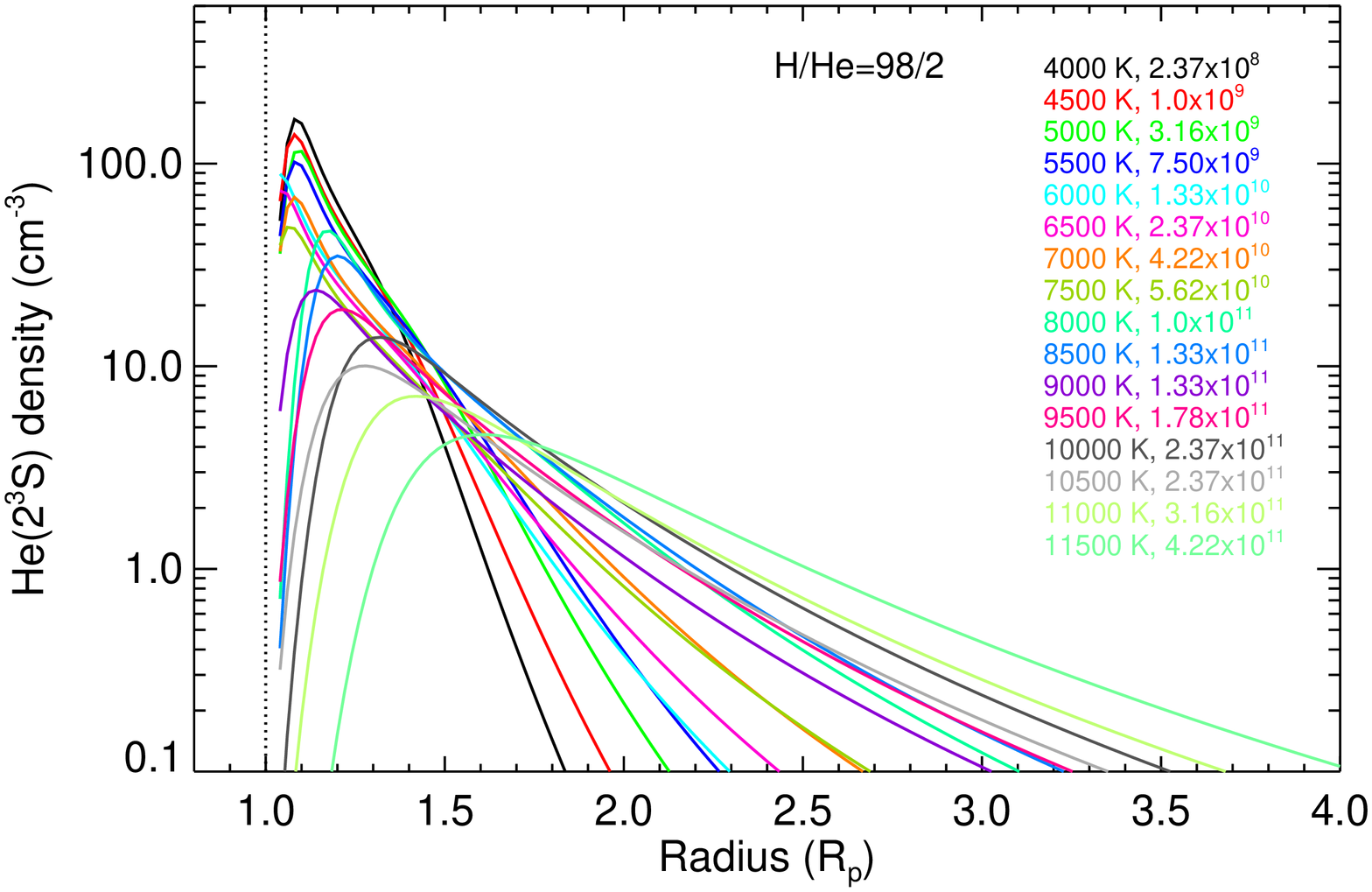}
\caption{\het\ concentration profiles that best fit the measured absorption (see Figs.~\ref{chi2} and \ref{chi2_2}). Shown is a selection of the profiles along the black, cyan. and orange lines in Fig.~\ref{chi2_2}, covering the whole  range of temperatures and mass-loss rates of those figures. Panels from top to bottom are for H/He ratios of 90/10, 95/5, and 98/2, respectively.} 
\label{he3_candidates} 
\end{figure}

Subsequently, we computed the spectral transmission for each of the helium triplet profiles.  
In these radiative transfer calculations, we included the temperature and the escape velocity profile $\ve(r)$ (see Eq.~\ref{eq:velocity_profile}) from the model. 
No turbulence term was included, as most of the absorption comes from radii below the Roche limit. Nevertheless, we performed a test for the case of Fig.~\ref{absorption} by including this term in the broadening of the line and found a less precise fitting; that is, the modelled line profile is wider than the measured profile. 

A bulk Doppler blueshift of 1.8\,\kms\ was included in the absorption calculation, as suggested by the observations.  The additional absorption observed at around --13\,\kms\ (the magenta curve in Fig.~\ref{absorption}) has not been included in these calculations (e.g. in the computation of $\chi^2$, see below). The reason is that this absorption seems to emerge from very high altitudes and possibly from material already ejected by the planet, but we are interested in determining the temperature and mass-loss rates of the bulk atmosphere. 

The absorption spectra were computed at a very high resolution (wavelength step of 10$^{-7}$\,\AA) and then convolved with the CARMENES line spread function. 
We performed the integration over the radius, $r$ in Fig.~\ref{sketch}, from the lower boundary condition of the He triplet model (1.04\,\rp) up to the Roche lobe boundary, located at 4.22~\rp\ \citep{Salz2016}. 

In order to analyse the extent to which absorption contributes beyond that limit, we also computed the absorption including the atmosphere extended up to 10\,\rp. The results show that for $\dot M$ smaller than 10$^{11}$\,\gs\,the additional absorption for radii beyond that boundary is very small; in particular, for $\dot M$ of $\sim$3.7$\times\,10^{9}$\,\gs\ at a temperature of 6000\,K (Fig.~\ref{absorption}) it is negligible (below 0.3\%). This result is consistent with the light curve shown in Fig.~5 in \cite{Alonso2019} which does not show signs of relevant ingress or egress absorption.

We compared the synthetic spectra to the measured one (see an example in Fig.~\ref{absorption}) and calculated the $\chi^2$ in the spectral interval of 10829.9 to 10831.5\,\AA\ (avoiding the --13\,\kms\ component) for each of them. 
In this way, we obtained all possible pairs of $\dot M$ and $T$ which are compatible with the measured absorption. 

The results of the $\chi^2$ computations for H/He of 90/10, our nominal case, are shown in Fig.~\ref{chi2}. The white curve highlights the best-fitting simulations. According to these results, the measured absorption is consistent with pairs of (T, \mlr) ranging from  (4625\,K,10$^{8}$\,\gs) to (11500\,K, 10$^{11}$\,\gs). The relationship between temperature and $\dot M$ is not linear, being steeper at lower $T$ and $\dot M$ values, and flatter at high temperatures.

In order to better understand the escaping atmosphere we plotted a selection of the \het\ densities  in Fig.~\ref{he3_candidates}
(top panel for an H/He ratio of 90/10) that fit the observed absorption spectrum and cover the full range of temperatures and $\dot M$. This figure shows that the majority of the \het\ profiles peak at very short distances, in the range of  1.04--1.10\,\rp. The few profiles with peaks at distances greater than 1.30\,\rp\ are more extended and generally correspond to higher temperatures and larger mass-loss rates. Another relevant feature of the atmospheres that best fit the measured absorption is that they all have a very similar averaged mean molecular weight, close to  0.69\,g\,mole$^{-1}$. This value corresponds to a partially ionised thermosphere, between the fully ionised ($\bar\mu$=0.65) and the neutral ($\bar\mu$=1.3) thermospheres.

\begin{figure}
\includegraphics[angle=90.0, width=1.0\columnwidth]{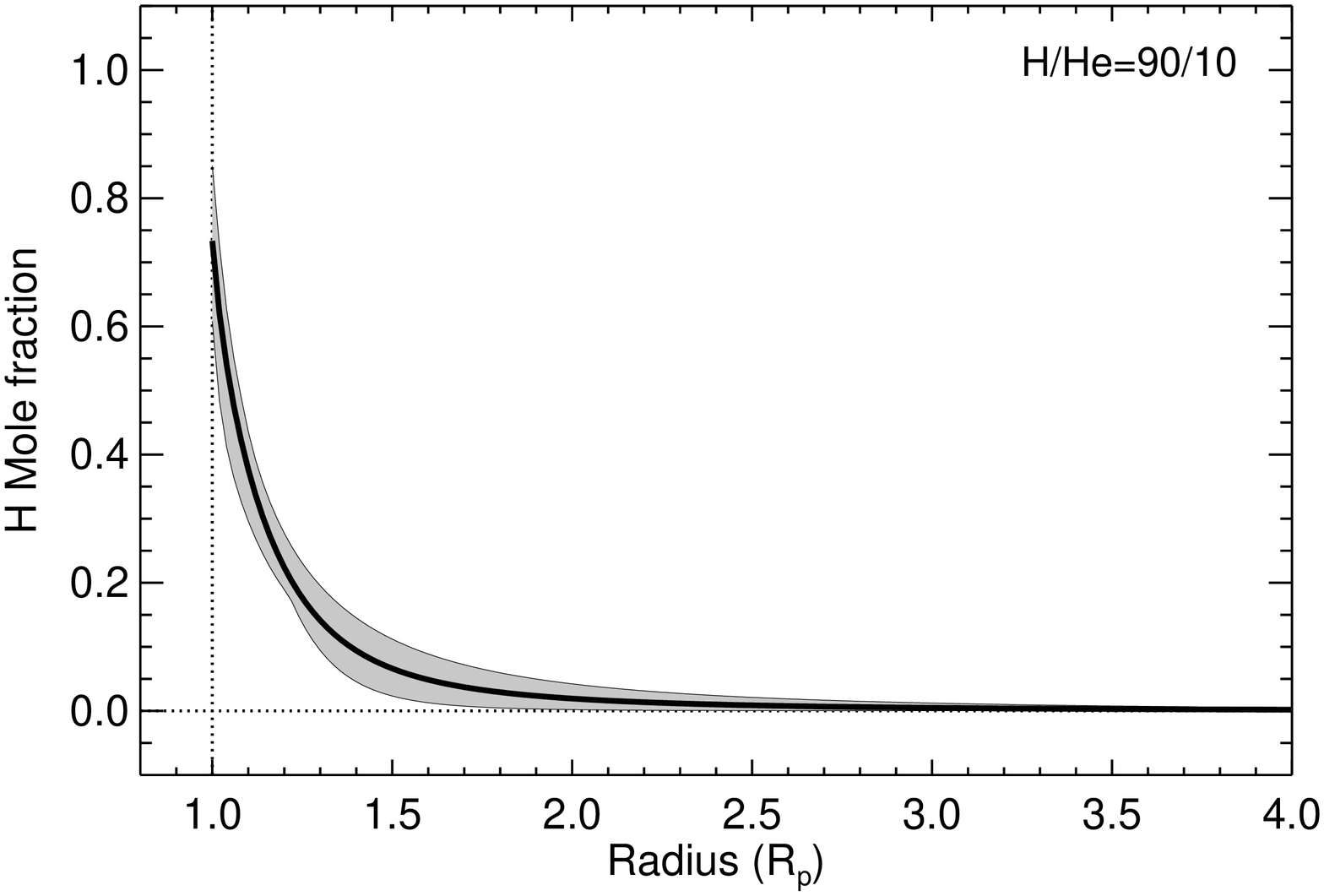}
\includegraphics[angle=90.0, width=1.0\columnwidth]{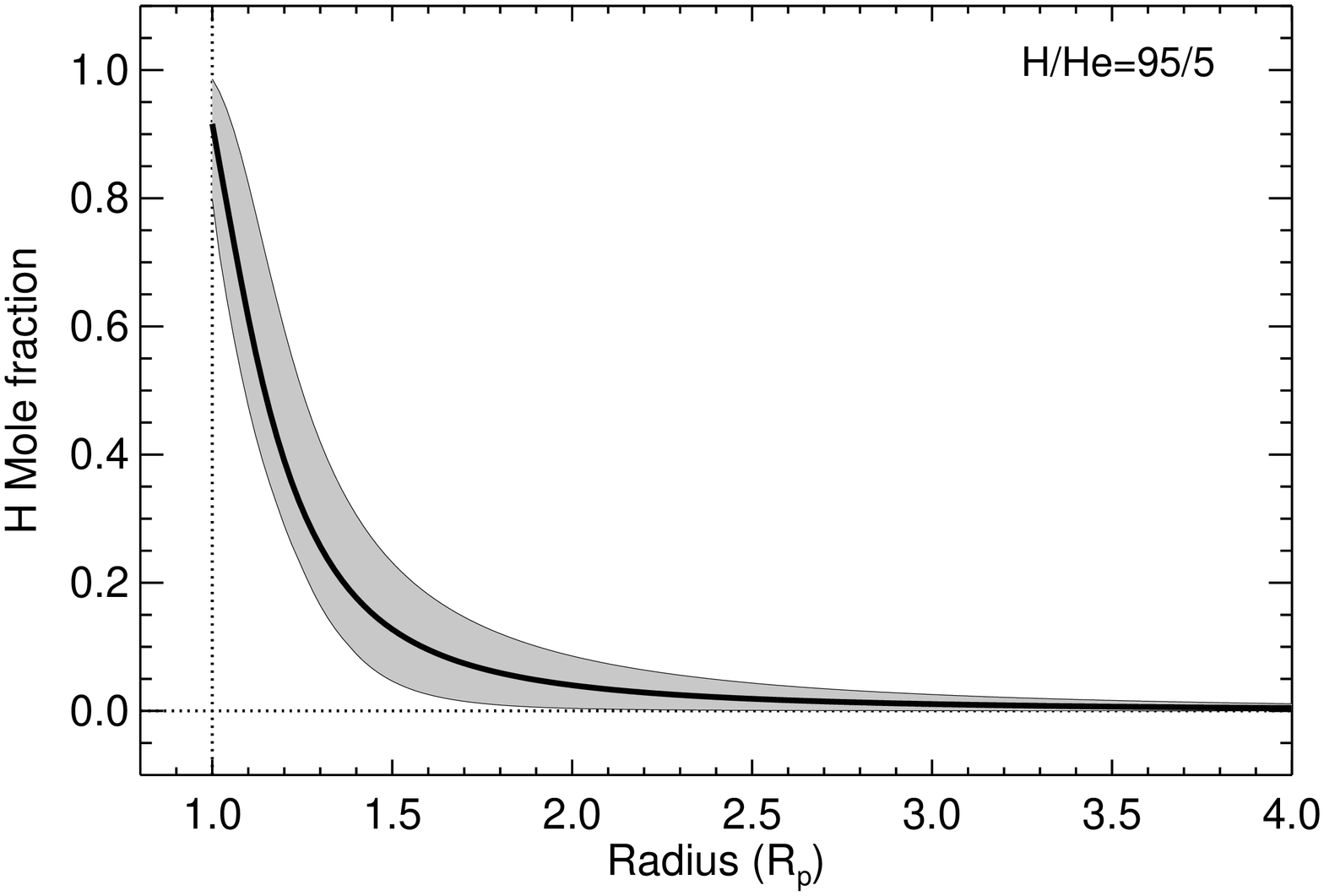}
\includegraphics[angle=90.0, width=1.0\columnwidth]{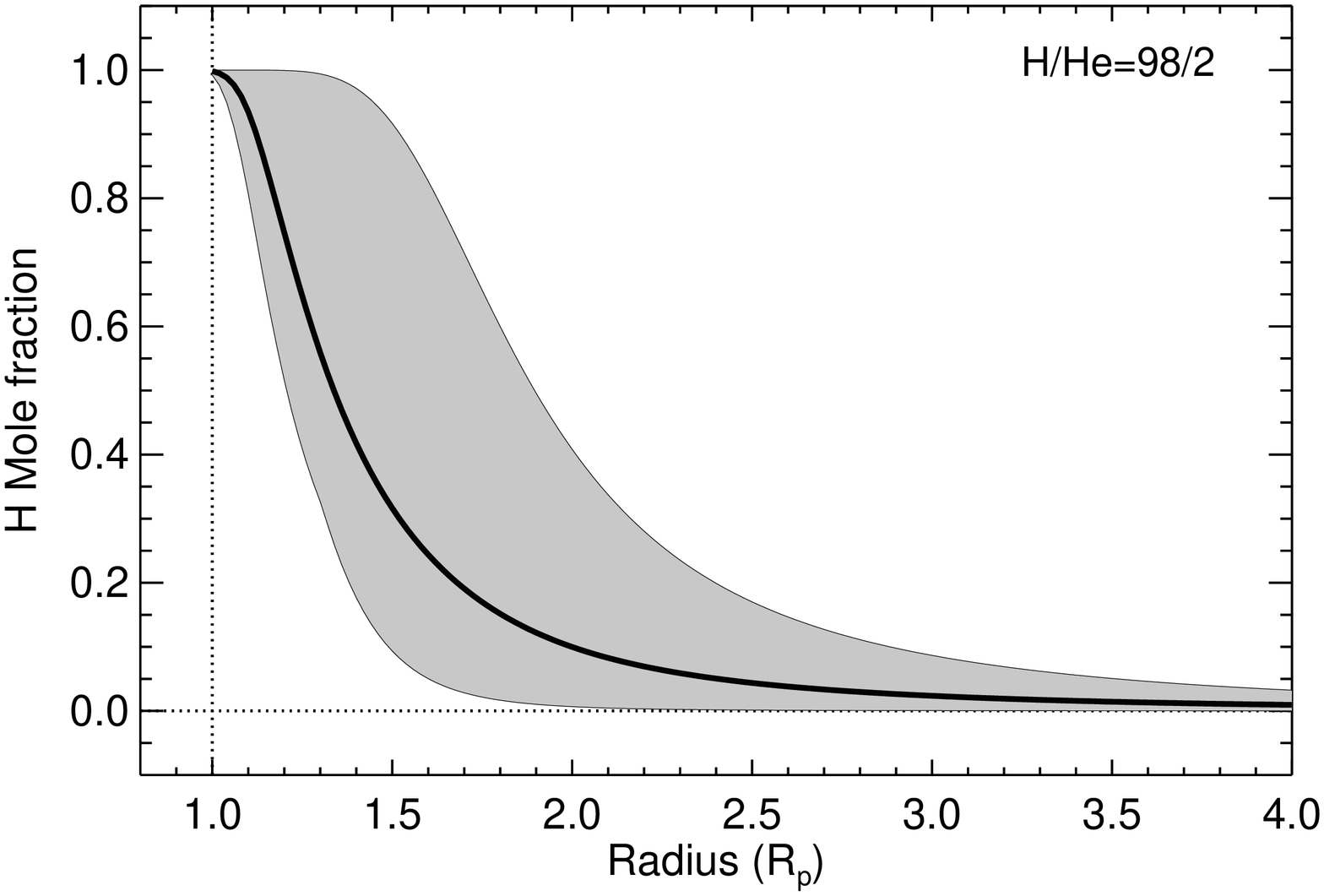}
\caption{Range of the hydrogen molar fraction profiles resulting from the fit of the measured absorption (see Figs.~\ref{chi2} and \ref{chi2_2}) for three H/He ratios, covering the whole temperature and mass-loss rate ranges of those figures. The solid thicker lines are the mean profiles.}
\label{h_candidates} 
\end{figure}

As the atmospheric model provides [H$^+$], we also analysed [H]/[H$^+$] for the different atmospheres that fit the  observed \het\ absorption. In Fig.~\ref{h_candidates} we show the H mole fraction profiles for the same atmospheres as shown in Fig.~\ref{he3_candidates}. We see that for an H/He of 90/10 (top panel) most of the atmospheres reproducing the measured \het\ absorption have a rather sharp [H]/[H$^+$] transition region occurring at altitudes ranging from about 1.04 to 1.09\,\rp. That is, 
the measured \het\ spectrum suggests that the atmospheric hydrogen of \hd20 is fully ionised at altitudes above $\sim$1.6\,\rp, which is in agreement with the mean molecular weight of $\sim$0.69\,g\,mole$^{-1}$ mentioned above.  

\begin{figure}
\includegraphics[angle=90.0, width=1.0\columnwidth]{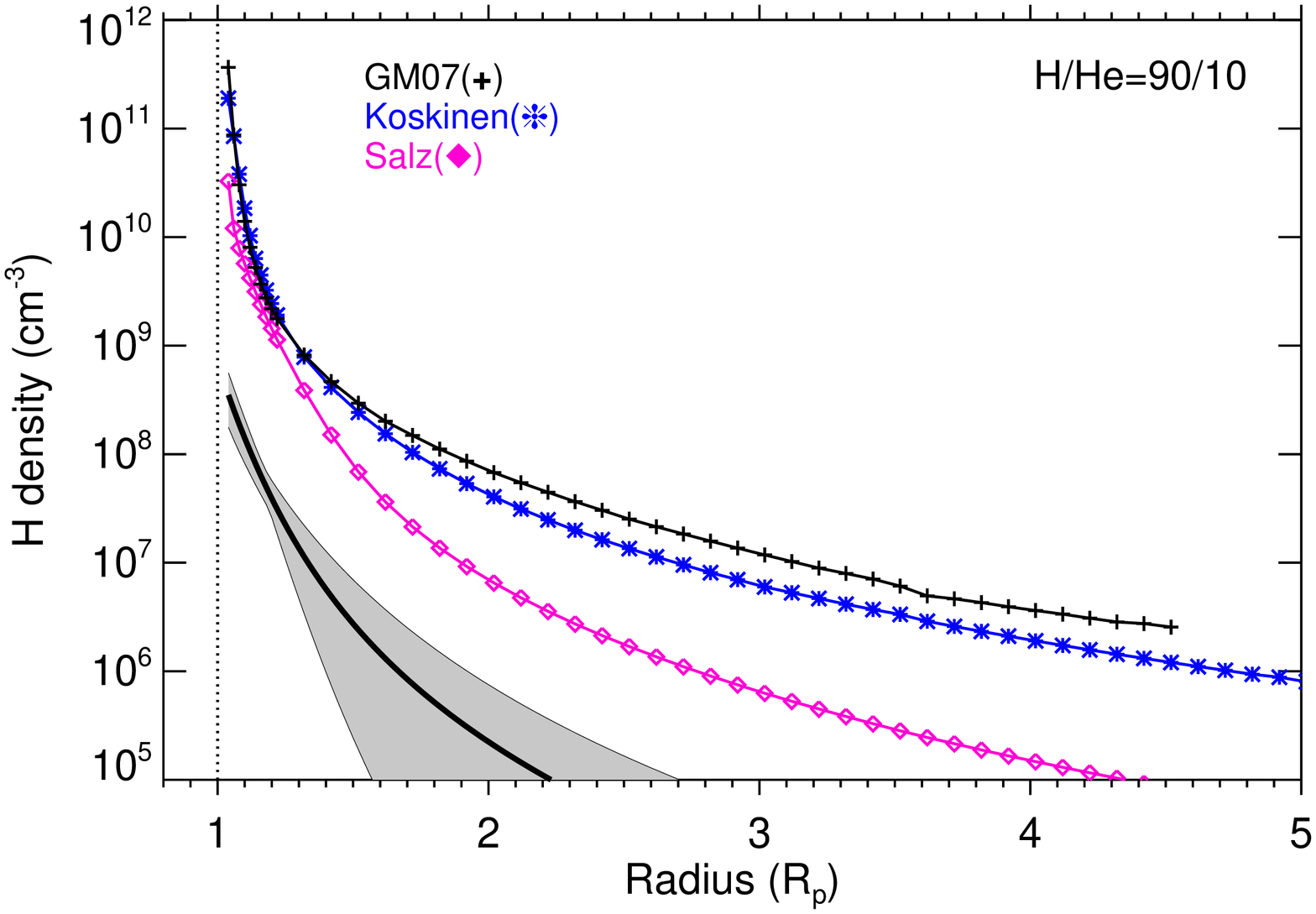}
\includegraphics[angle=90.0, width=1.0\columnwidth]{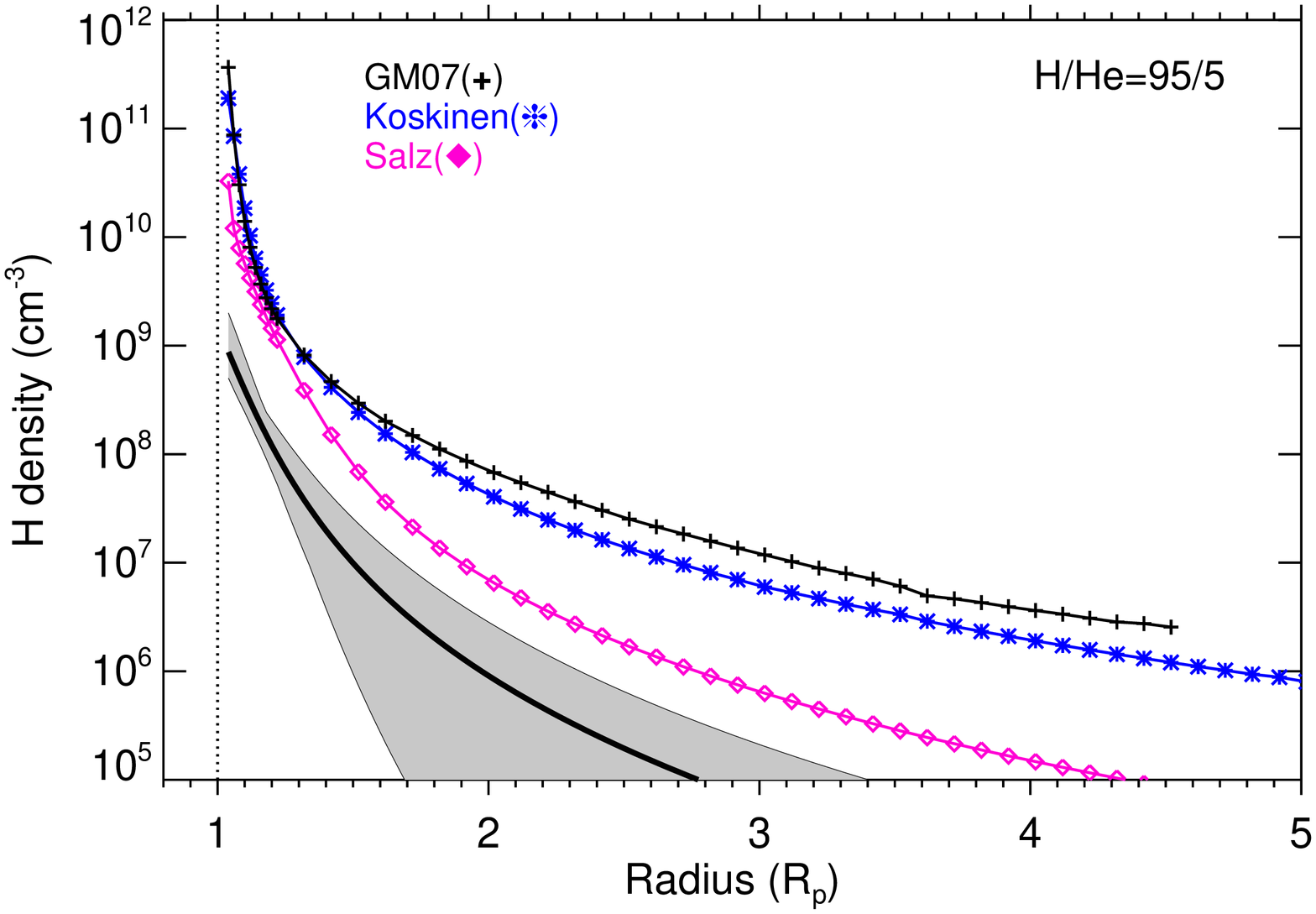}
\includegraphics[angle=90.0, width=1.0\columnwidth]{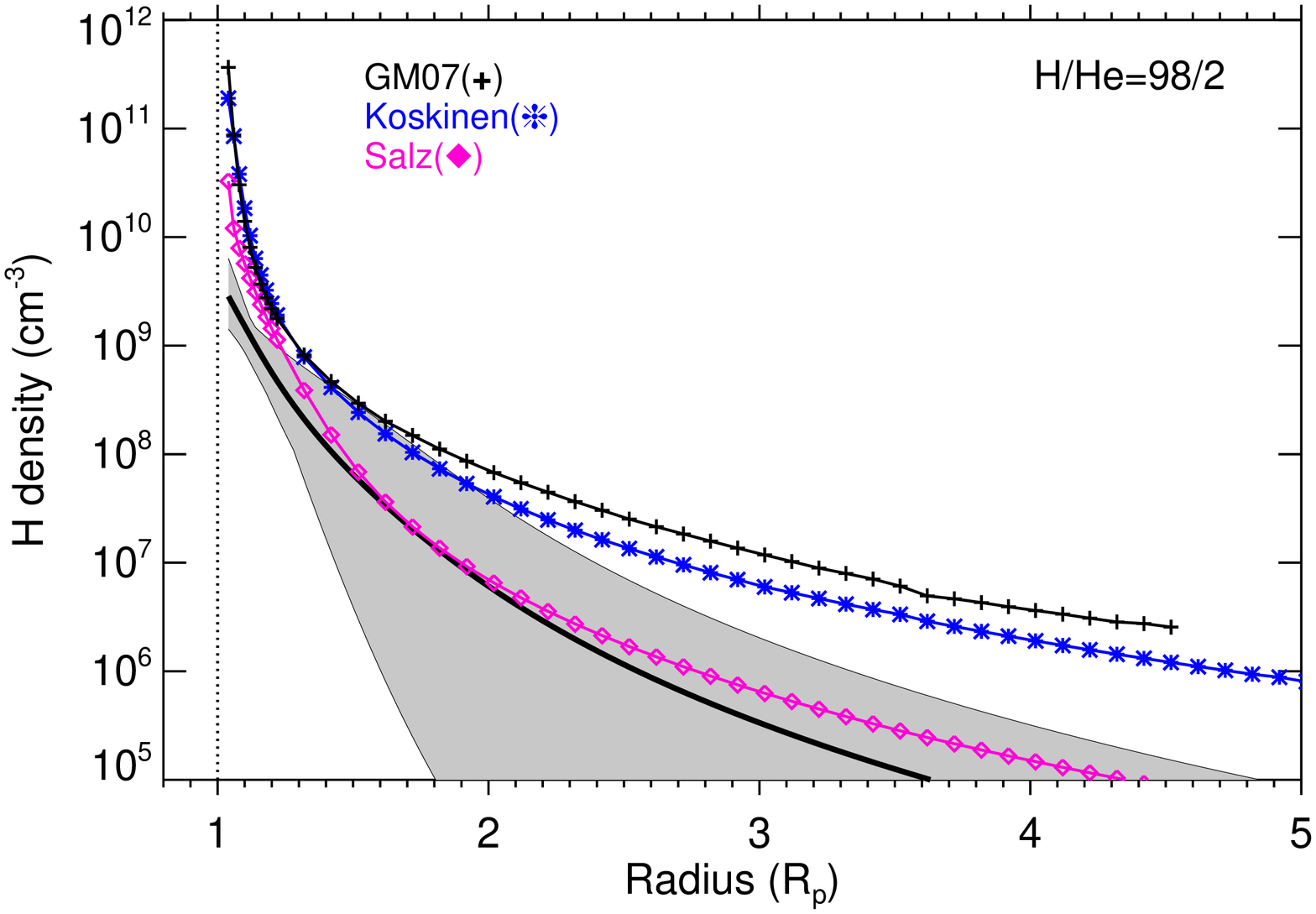}
\caption{Range of the neutral hydrogen concentration profiles resulting from the fit of the measured absorption (see Figs.~\ref{chi2} and \ref{chi2_2}) for three H/He ratios, covering the whole temperature and mass-loss rate ranges of those figures. The solid thicker lines are the mean profiles. The H density derived from \lya\ measurements reported by \cite{Garcia_munoz_2007}, \cite{Koskinen2013a}, and \cite{Salz2016} is also shown.}
\label{hden_candidates} 
\end{figure}

\begin{figure}
\includegraphics[angle=90, width=1.0\columnwidth]{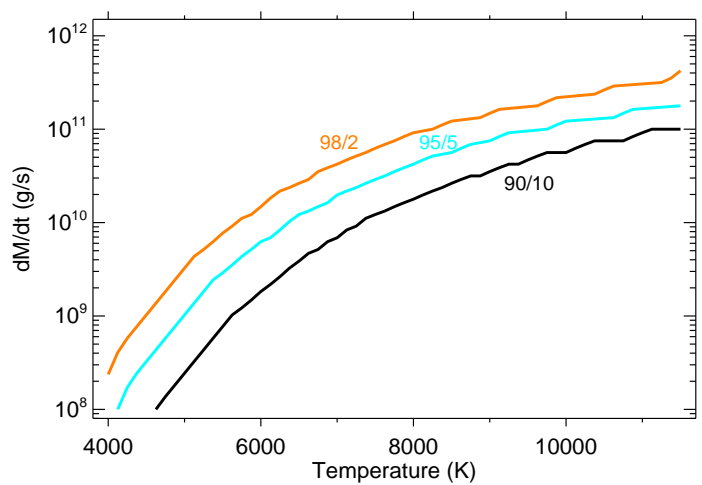} 
\caption{Lines corresponding to the (T, \mlr) pairs of the model that best fit the helium triplet absorption as measured by CARMENES. The black curve shows the results for an H/He ratio of 90/10 (e.g. white curve in Fig.\,\ref{chi2}). The cyan and orange lines correspond to the best fit obtained when assuming H/He ratios of 95/5 and 98/2, respectively.}  
\label{chi2_2}
\end{figure}

\begin{figure}
\includegraphics[angle=90.0, width=1.0\columnwidth]{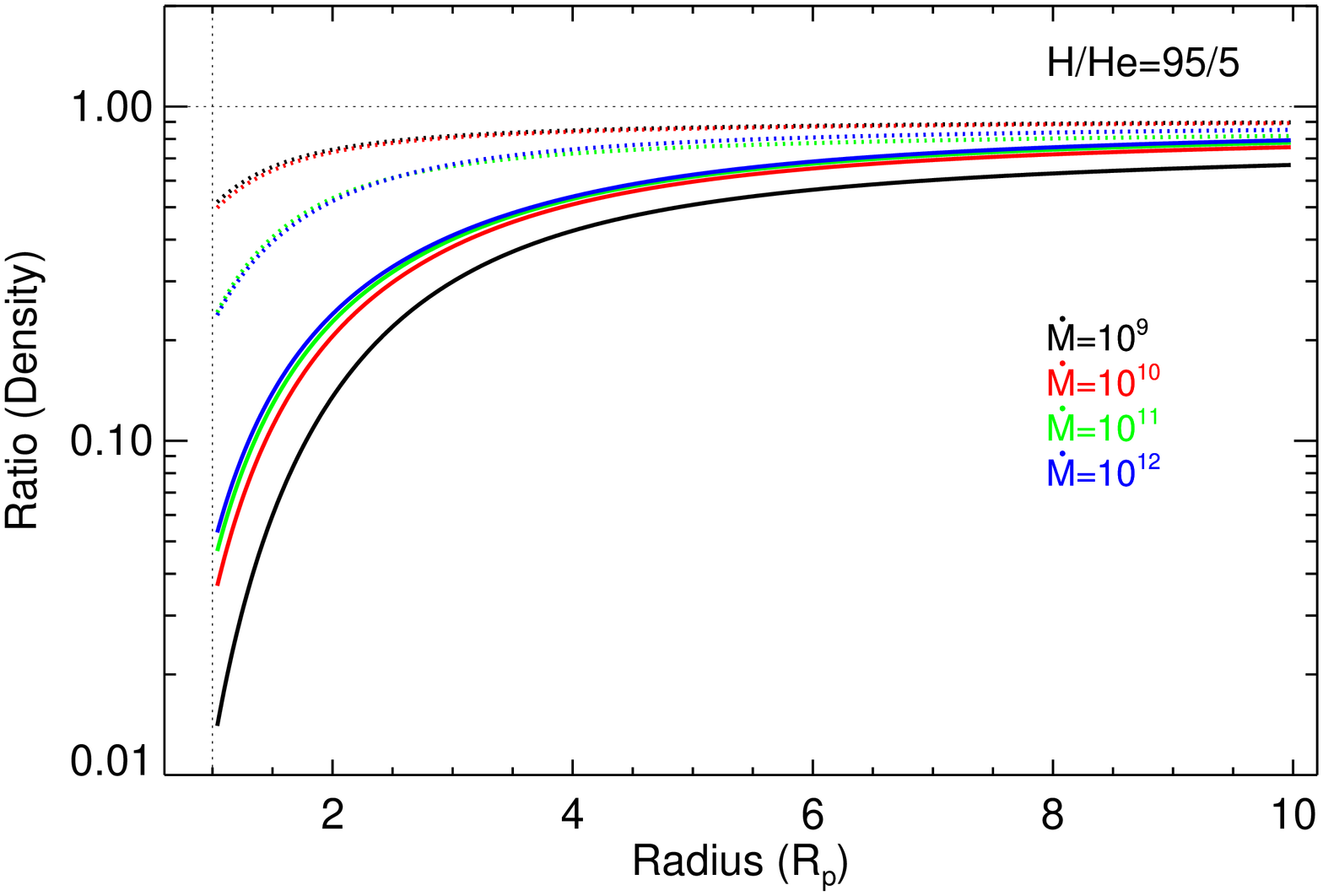}
\includegraphics[angle=90.0, width=1.0\columnwidth]{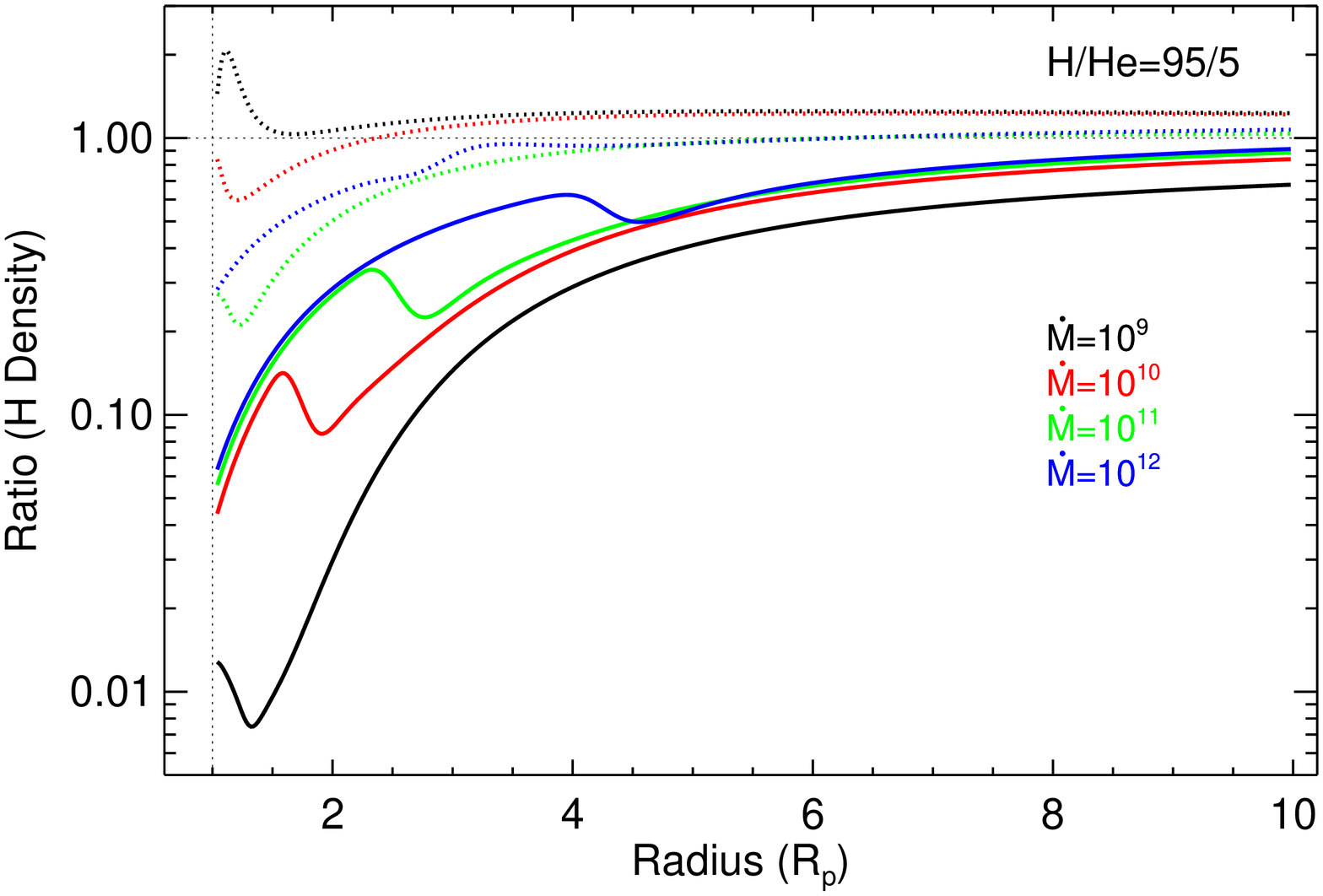}
\includegraphics[angle=90.0, width=1.0\columnwidth]{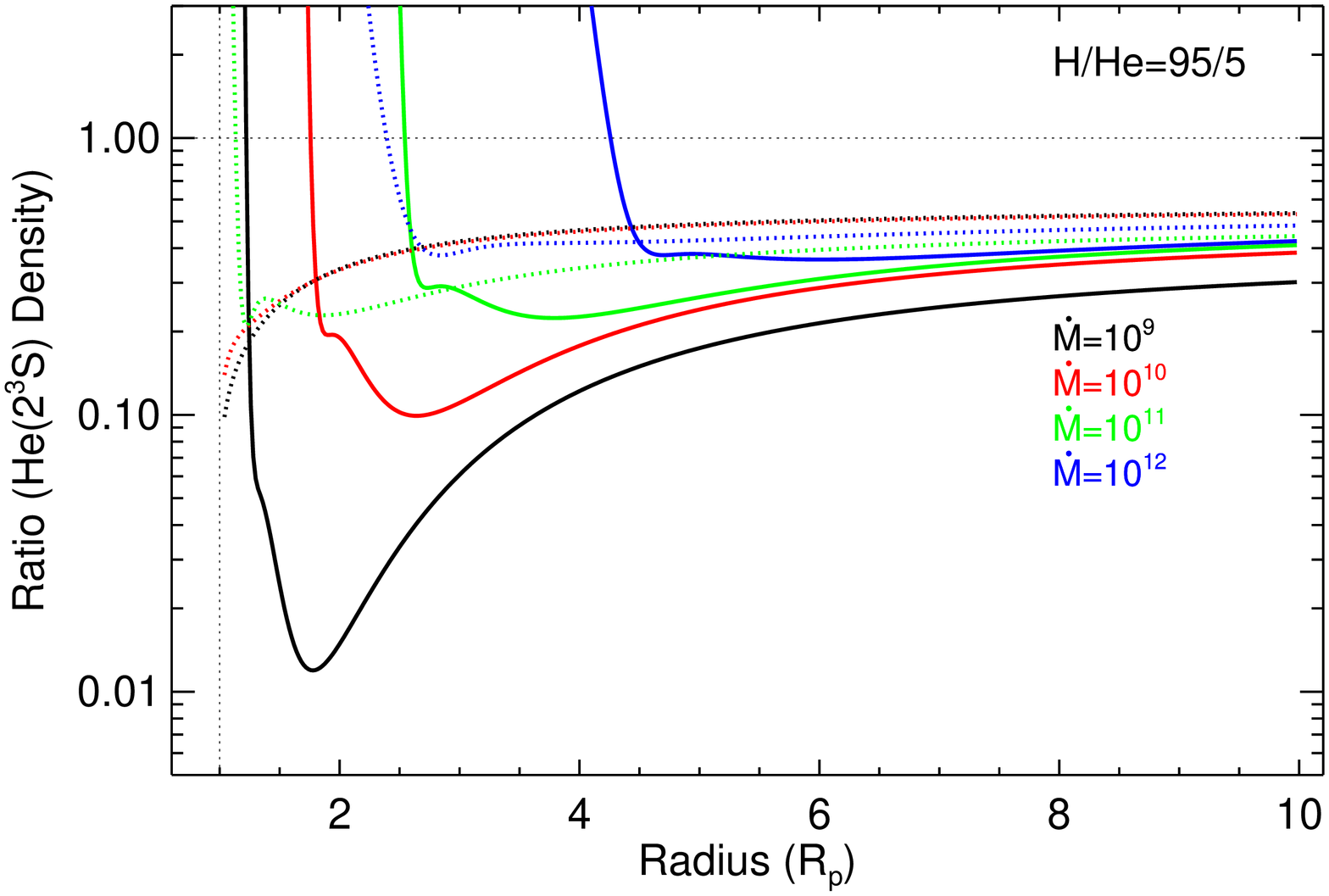}
\caption{Ratios of the total atmospheric mass density (top panel), the neutral H number density (middle panel), and the \het\ density (bottom panel) for several mass-loss rates and temperatures of 5000\,K (solid) and 9000\,K (dotted) for an H/He ratio of 95/5 relative to the densities for an H/He ratio of 90/10.} \label{den_ratio} 
\end{figure}

\subsection{Sensitivity to the H/He ratio and comparison with previous H densities}\label{h_he_ratio}

We further explored whether or not we can constrain the H concentration with the measured \het\ spectrum. The H abundance in \hd20\ has been derived from \lya\ absorption measurements in several studies 
\citep[see, e.g.][]{Garcia_munoz_2007,Koskinen2013a,Salz2016}. We then compared our derived H density from the \het\ fit with the ones reported in these latter studies. We found that, in general, our derived H densities for the usual H/He ratio of 90/10 (see Fig.~\ref{hden_candidates}, top panel) are significantly lower at all altitudes. This prompts us to consider other H/He densities in order to ascertain whether or not they can explain these discrepancies. Therefore, we repeated all the calculations above for the larger H/He ratios of 95/5 and 98/2. 
There are several processes that could lead to a larger H/He   than the canonical value of 90/10. For example, as He is not chemically active, depending on the location of the He homopause, its abundance can be lower than its canonical value at lower altitudes due to diffusive separation \cite[see, Fig. 14 in][]{Moses2005}. Another possible process of depletion He in the upper atmosphere of a giant planet is the He sequestration in the interior of the planet due to the formation of an H--He immiscibility layer \citep{Salpeter1973,Stevenson1975, Stevenson1980, Wilson2010}.

The most noticeable effect of increasing the H fraction (and thus decreasing the He abundance) is a global decrease of the 
thermospheric mass density. Therefore the thermosphere is lighter and, for given conditions of temperature and mass-loss rate, it is more expanded. Figure~\ref{den_ratio} (top panel) shows the change in the \hd20\ density when assuming an H/He ratio of 95/5. Here we observe large changes in the density of close to two orders of magnitude for lower temperatures at low distances. The H density also shows large decreases (Fig.~\ref{den_ratio}, middle panel), following the total density very closely. We also see the effects of absorption of the stellar radiation mainly for lower temperatures; that is, a change in the slope of the H density ratio at distances close to 1, 2, 3, and 4.5\,\rp\ (for \mlr\ of 10$^9$ to 10$^{12}$\,\gs) in response to the penetration of radiation to deeper altitudes for smaller H/He ratios. 

As a consequence of the lower total density, and lower H density, for a larger H/He ratio, the helium triplet density also mainly follows the behaviour of H, and therefore we generally obtain smaller \het\ densities (Fig.~\ref{den_ratio}, bottom panel). It is also worth noting that the peak of the \het\ density moves to  slightly lower altitudes for larger H/He.

The results of fitting the measured \het\ spectrum when using H/He of 95/5 and 98/2 in the model are shown in the corresponding figures: Figure~\ref{chi2_2} (cyan and orange lines) for the constrained (\mlr, T) pairs; Fig.~\ref{he3_candidates} (middle and bottom panels) for the \het\ density; and in Figs.~\ref{h_candidates} and \ref{hden_candidates} (middle and bottom panels) for the H mole fraction and H density, respectively. 

Regarding the (\mlr, T) relationship, this curve moves to colder temperatures and larger mass-loss rates for higher H/He (see  Fig.~\ref{chi2_2}), which is the consequence of a smaller \het\ density. In order to fit the measured spectrum we need to lower $T$ or increase \mlr  \ in order to increase the \het\ concentration (see Fig.~\ref{fig_He3density}). For the relatively high H/He ratio of 98/2, the measured spectrum implies relatively low temperatures and high mass-loss rates.

Figure\,\ref{he3_candidates} (middle and bottom panels) shows the \het\ densities for the higher H/He of 95/5 and 98/2. In comparison to the result obtained for the commonly used H/He ratio of 90/10, the concentrations peak at generally lower values. These peaks are located at larger radii, and the profile shapes are wider.
The chosen H/He ratio used in the fitting of the \het\ absorption also dictates the ionisation level of the atmosphere. Figure~\ref{h_candidates} shows that the atmosphere of \hd20 is ionised at higher altitudes for larger H/He.

By inspecting the results obtained for the H density for the three H/He ratios considered here (see Fig.~\ref{hden_candidates}), it seems that the derived H densities that agree better with those derived from previous \lya\ measurements correspond to the 98/2 ratio (lower panel).

\subsection{Comparisons of temperatures and mass-loss rates to those of previous works}\label{comparison}

Several models with different approaches and assumptions have been developed for studying the processes that drive the escape in \hd20. A summary is listed in Table\,\ref{table.authors} for comparison. Before discussing them, we should recall that our mass-loss rates refer to the total mass of hydrogen and helium, and that they were computed at the substellar location instead of the globally averaged values reported by some authors, the latter being four times smaller. When necessary, we translate globally averaged  to substellar values.

One of the approaches for escape modelling is to consider a particle model, which focuses on the exosphere or the unbound region. With the aim of interpreting \hd20 observations of H\,{\sc i} Ly$\alpha$, \cite{VidalMadjar2003} built a particle model where the atmosphere is composed  of atomic hydrogen only. These latter authors considered the stellar radiation pressure and the planetary and stellar gravity. Thereby, they determined a lower limit for the substellar mass-loss rate of 0.4 $\times\,10^{11}$\,\gs. 
This value, when compared to those derived here (we consider that our case of an H/He ratio of  98/2 gives the better overall agreement for the \het\ and Ly$\alpha$ absorption; see Fig.~\ref{chi2_2}, orange curve) would implies a temperature of about 7000\,K, which seems plausible although generally lower than those calculated in different models \cite[e.g.][]{Salz2016}. Higher temperatures lead to stronger mass-loss rates, which is consistent with  \citeauthor{VidalMadjar2003} since they provide a lower limit. 

Similarly, \cite{Bourrier_2013} developed a 3D model for an atomic hydrogen exosphere including stellar radiation pressure. These latter authors suggested a substellar mass-loss rate range of (0.04\,--\,4.0)\,$\times\,10^{11}$\gs. This wide range of mass-loss rates is compatible with our results but does not significantly constrain the temperature range because, according to Fig.~\ref{chi2_2} (orange curve), the temperature varies in a wide range: 5125--11500\,K.

Other studies using hydrodynamic models have focused on thermospheric rather than exospheric simulations. In these approaches, several 1D models with spherical geometry have been developed. For example, \cite{Tian2005}, in their case for an atmosphere of atomic hydrogen and a heating efficiency of $\eta$\,=\,0.6, predicted a maximum substellar mass-loss rate of 2.4\,$\times\,10^{11}$\,\gs\ for a temperature profile that grows monotonously with altitude from 800\,K at 1\,\rp\ to $\sim$20000\,K at 3\,\rp. 
For this mass-loss rate, we derive from Fig.~\ref{chi2_2} a temperature close to 10500\,K, which is significantly lower than most of the values predicted by \cite{Tian2005}.
Nevertheless, \cite{Shematovich_2014} concluded (see discussion below) that heating efficiencies 
smaller than 0.2 probably yield more accurate mass-loss rates. Thus, the results from \cite{Tian2005} probably represent an overestimation of temperature.

Also, using atomic hydrogen thermospheric models and assuming heating efficiencies of 0.6 and 0.1, \cite{Penz_2008} estimated substellar mass-loss rates of 1.4\,$\times\,10^{11}$ and 0.52\,$\times\,10^{11}$\,\gs, and maximum temperatures of 9500\,K and 6300\,K located near 1.5\,\rp, respectively.
For these mass-loss rates, we derive from Fig.~\ref{chi2_2} temperatures close to 9100\,K (for their $\eta$ = 0.6), and of 7200\,K (for their $\eta$ = 0.1), which is comparable for $\eta$=0.6 but slightly higher than that of the latter authors for $\eta$=0.1. We note, however, that they include only hydrogen. Thus, if we were to consider 
a lower He abundance in our analysis, our results would be in better agreement with theirs.

\cite{Garcia_munoz_2007} studied the hydrodynamic escape of \hd20 for different atmospheric compositions. The most relevant case is that with a complete chemical scheme (H, He, C, O, N and D) and an H/He ratio of $\approx$91/9 (his DIV1 case), for which this latter author estimated a mass-loss rate of 5.0\,$\times\,10^{11}$\,\gs. Furthermore, the maximum temperature found by this latter author is 12500\,K at radii of 1.2\,\rp.
For his mass-loss rate we derive from Fig.~\ref{chi2_2} (orange curve) a temperature of 11500\,K or even larger, which is in good agreement with that computed by \citeauthor{Garcia_munoz_2007}.
 
\cite{Koskinen2013a} carried out simulations for a thermosphere composed of H, He, and heavier elements (C, O, N, Si, Mg, Na, K and S). For their C2 case, which is the most representative, these latter authors obtained a mass-loss rate of 1.6\,$\times\,10^{11}$\,\gs\, with a heating efficiency of 0.44, and a maximum temperature of 12000\,K at a distance of 1.5\,\rp. For this mass-loss rate we obtain a temperature of about 9100\,K, 
which is significantly lower than theirs.

\begin{table*}
\centering
\caption{\label{table.authors}Comparison of temperature and mass-loss rates with other works.}
\begin{tabular}{c c c c c c c c l l} 
   \hline
   \hline
   \noalign{\smallskip}
$F_{\rm XUV}$ & H/He & $R_{\rm XUV}$ & K & $\eta$$^{(a)}$ & $\dot M$$^{(b)}$ & T$^{(c)}_{{\rm max}}$ & T$^{(d)}_{\rm This ~work}$  & Model$^{(e)}$ & References \\
(erg cm$^{-2}$ s$^{-1}$) & & (R$_{\rm p}$) & & & ($\times 10^{11}$\,g s$^{-1}$ ) & ($\times 10^{3}$\,K) & ($\times 10^{3}$\,K) & &\\
  \noalign{\smallskip}\hline\noalign{\smallskip}
... & 100/0 & &  &  & > 0.4 $^{(f)}$ &  & > 7.0 & 3DP & (1)     \\
 (i) & 100/0  & & &  &  0.04--4.0 & & 5.1--11.5 &  3DP & (2)    \\
(h) &  100/0  & &  & 0.6 & 2.4 & see text & 10.5 &  HD & (3)    \\

1976 & 100/0  & &  & 0.6 & 1.4 & 9.5 & 9.1 &  HD & (4)\\
1976 & 100/0  & &  & 0.1 & 0.52 & 6.3 & 7.2 &  HD & (4)\\
2900 & 91/9 & &  & see text & 5.0 & 12.5 & 11.5 &  HD & (5)     \\
1800 & ... & &  & 0.44 & 1.6 & 12.0 & 9.1 &  HD & (6)   \\
1148 & 90/10 & &  & see text & 0.74 & 9.1 &  7.8 &  HD & (7)    \\
  \noalign{\smallskip}\hline\noalign{\smallskip}
(h) & & 3.0 & 1.0 & 1.0 & < 40$^{(g)}$ &   & see text &  EL & (8)       \\
910 & & 1.0 & 0.65 & 1.0 & < 1.8$^{(g)}$ &   & < 9.8 &  EL & (9)        \\
1148 & & 1.25 & 1.0 & 0.21 & 0.74 &   &  7.8 &  EL & (10)       \\
2400 &  98/2 & 1.16-1.30$^{(j)}$ & 0.76 & 0.1-0.2$^{(k)}$ &  0.42--1.00 &  & 7.13--8.13 &   & This work   \\

\noalign{\smallskip}
\hline
\end{tabular}
\tablefoot{
\tablefoottext{a}{Averaged heating efficiencies. Particle models (3DP) do not include this parameter.}
\tablefoottext{b}{Substellar mass-loss rates. Globally averaged mass-loss rates were translated to substellar when necessary, multiplying by 4.}
\tablefoottext{c}{Maximum temperature. Given only for hydrodynamic (HD) models.}
\tablefoottext{d}{Temperature of our model for a 98/2 H/He abundance, corresponding to the \mlr\ of the compared model.}
\tablefoottext{e}{Type of model: 3D particle estimation (3DP), Hydrodynamic model (HD) and Energy-limited (EL).}
\tablefoottext{f}{Lower limit of mass-loss rate.}
\tablefoottext{g}{Upper limit of mass-loss rate.}
\tablefoottext{h}{Current solar XUV flux.}
\tablefoottext{i}{Three to four times the current solar XUV flux.}
\tablefoottext{j}{$R_{\rm XUV}$ range derived in this work for the assumed heating efficiency range (0.1-0.2).}
\tablefoottext{k}{Range taken from \citet{Shematovich_2014}.} 
}
\tablebib{
(1) \citet{VidalMadjar2003};
(2) \citet{Bourrier_2013};
(3) \citet{Tian2005};
(4) \citet{Penz_2008};
(5) \citet{Garcia_munoz_2007};
(6) \citet{Koskinen2013a};
(7) \citet{Salz_2015};
(8) \citet{Lammer_2003};
(9) \citet{san11};
(10) \citet{Salz2016};
}
\end{table*}

\begin{figure}
\includegraphics[angle=90.0, width=1.0\columnwidth]{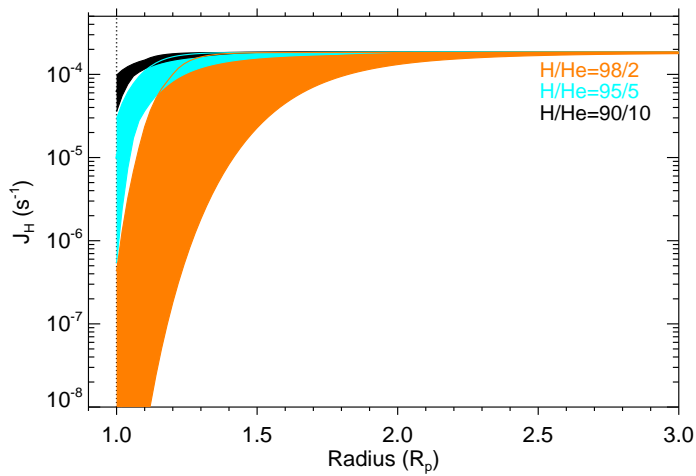}
\caption{Ranges of the photo-ionisation coefficient of H, $J_H$ profiles that fit the measured absorption (Fig.~\ref{chi2_2}) for the different H/He ratios.} 
\label{J_H} 
\end{figure}

For an atomic hydrogen and helium thermosphere of 90/10, \cite{Salz2016} estimated a mass-loss rate of 0.74\,$\times\,10^{11}$\,\gs\ with a maximum temperature of 9100\,K near 1.4\,\rp. For that mass-loss rate our \het\ absorption implies a temperature of about  7750\,K (Fig.~\ref{chi2_2}, orange curve, H/He of 98/2), which is significantly lower than the temperature of \citeauthor{Salz2016}.
It is interesting to note that \citeauthor{Salz2016} obtained an H density that underestimates the \lya\ observations. For an H/He ratio of 90/10, our H density derived from the \het\ absorption also underestimates the \lya\ absorption (see top panel in Fig.~\ref{hden_candidates}). 
In order to fit the \lya\ observations, \citeauthor{Salz2016} supplied the required extra H using an H-only (no helium) model. In our case, to obtain a good fit to both the He triplet and \lya\ absorption, 
we also need to decrease the H/He ratio; suggesting that a value of 98/2 could be enough.

Overall, we observe that although the ranges of mass-loss rates and temperatures in the literature are rather broad, they generally agree well with the constrained \mlr(T) curve found in this work (orange curve in Fig.~\ref{chi2_2}). Only the values derived by \cite{Tian2005} 
and \cite{Koskinen2013a}, both larger than ours, fall relatively far from the derived constraints.

\subsection{Energy-limited escape models}

Estimates of the evaporation mass-loss rate of \hd20 have also been obtained by using the energy-limited approximation \citep{Watson_1981,Erkaev_2007,Lammer_2009}. 
The main assumption of this model is that the escape is limited by the $F_{\rm XUV}$, and it is useful for constraining the atmospheric mass-loss rate or the heating efficiency. With this approach, the substellar mass-loss rate, $\dot M_{EL}$, can be written as 
\begin{align}
\dot M_{EL} = \frac {4 \pi\, R_{p}\, R_{XUV}^{2}\, \eta\, F_{XUV}} {G\,K(\xi)\, M_{P}},
\label{eq:energy_lim}
\end{align}
where $R_{\rm XUV}$ is the atmospheric expansion radius, defined as the altitude where the optical depth is unity \citep{Watson_1981}; 
$K(\xi) = 1 - 1.5\,\xi + 0.5\,\xi^{3}$ is the potential energy reduction factor, with $\xi = \left(M_{P}/M_{\star} \right)^{1/3}\left(a/R_{P}\right)$. 
This mass-loss rate is sometimes corrected by a factor that compensates for the underestimated kinetic and thermal energy gains. Here, however, we use uncorrected values.

\begin{figure}
\includegraphics[angle=90.0, width=1.0\columnwidth]{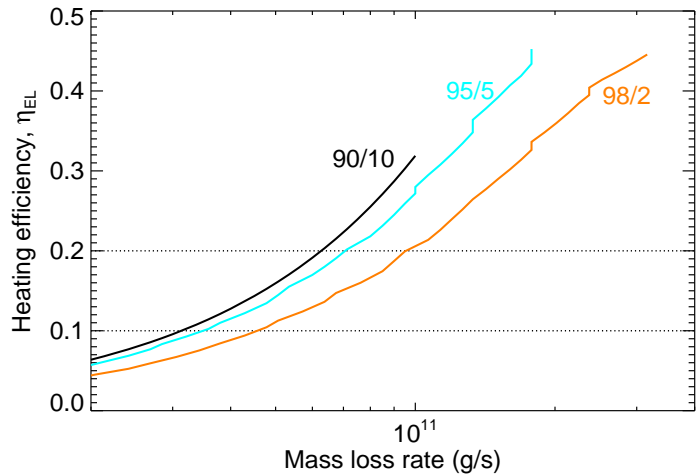}
\caption{Heating efficiency derived from the \het\ absorption measurements (see Fig.~\ref{chi2_2}) under the energy-limited escape approximation (see Eq.~\ref{eq:energy_lim}) for H/He of 90/10, 95/5, and 98/2. Dotted lines correspond to $\eta$=0.1 and 0.2.}
\label{eta} 
\end{figure}

We calculated the heating efficiency, $\eta_{EL}$, resulting from our \het\ analysis using the aforementioned approach. To that end, we included in Eq.~(\ref{eq:energy_lim}) the XUV stellar flux that we used, $F_{\rm XUV}$\,=\,2.4$\times$\,$10^3$\,erg\,cm$^{-1}$\,s$^{-1}$ (see Sect.\,\ref{atm_model}), $K$\,=\,0.76, and the $R_{\rm XUV}$ values resulting from the model fit to the \het\ absorption (see Fig.~\ref{J_H}). 
The resulting heating efficiencies are shown in Fig.~\ref{eta}.

As discussed above, \cite{Shematovich_2014} modelled the thermal-escape-related heating efficiency for \hd20 and concluded that $\eta$ values smaller than 0.2 produce more realistic mass-loss rates.
If we impose $\eta$=0.1--0.2 and use the energy limited approach, our results of Fig.~\ref{eta} limit the substellar mass-loss rate to the range of  
(0.42 -- 1.00)$\times10^{11}$\,\gs\ (on the basis that the H/He ratio of 98/2 gives the best fit to the H density), which, together with the results of Fig.~\ref{chi2_2}, leads to a temperature range of 7125--8125\,K (see Table\,\ref{table.authors}, last row).

Depending on the values used for the different parameters in Eq.~(\ref{eq:energy_lim}), we find a wide range of mass-loss rate and heating efficiency estimates for \hd20. For instance, \cite{Salz2016} estimated a heating efficiency of 0.21 using a mass-loss rate of 0.74$\times 10^{11}$\,\gs. These values are in good agreement with those obtained in this work.

This approach has also been used to provide upper limits to the mass-loss rate, that is, when all radiation is thermalised ($\eta$\,=\,1).  Thus, \cite{Lammer_2003} estimated an upper limit for the mass-loss rate of $\approx$\,40\,$\times 10^{11}$\,\gs\  using a calculated expansion radius of $\approx$\,3\,$R_{\rm p}$ and assuming $K$\,=\,1.0. This value is, in effect, a very high upper limit as it is much larger than those derived in this work for reasonable temperatures ($<$11500\,K) of 
$\sim$\,4\,$\times 10^{11}$\,\gs\ (see Fig.~\ref{chi2_2}). 

\cite{san11} estimated an upper limit of 1.8\,$\times 10^{11}$\,\gs\ assuming $K$ = 0.65, $R_{\rm XUV}$ equal to unity, and $F_{\rm XUV}$= 910\,erg\,cm$^{-1}$\,s$^{-1}$. This upper limit 
agrees with our results. We note that these latter authors used a simplified value for $R_{\rm XUV}$ and an $F_{\rm XUV}$ flux that is 2.6 times lower than ours.    

\section{Conclusions} \label{conclusions}

We present observational constraints on the thermosphere of \hd20 obtained from an analysis of the mid-transit \het\ spectral absorption measurements reported in \cite{Alonso2019}. Under the assumption of an upper atmosphere in blow-off escape, we modelled its thermosphere using a 1D hydrodynamic model with spherical symmetry, coupled with a non-LTE model for the population of the \het\ state. The model provides, among other quantities, profiles of the \het\ concentrations, which were included into a radiative transfer model to calculate \het\ absorption spectra. By comparing these spectra to the measured \het\ spectrum of \cite{Alonso2019}, we constrained the mass-loss rate, temperature, \het\ densities, and the degree of ionisation in the upper atmosphere of \hd20. We also compared the derived H densities with previous profiles derived from the \lya\ absorption in order to further constrain the H density as well as the H/He ratio.

One of the main results we found is the close relationship between mass-loss rate and temperature for a wide range of these parameters (see Fig.~\ref{chi2}). Additionally, we found that this relationship changes with the assumed H/He ratio (see Fig.~\ref{chi2_2}). For a given temperature, the lighter the atmosphere, the stronger the mass-loss rate; and, for a fixed mass-loss rate, the lighter the atmosphere, the cooler the atmosphere. The H/He degeneracy was partially constrained by comparing with previous H density derived from \lya\ absorption measurements. Globally, we found that an H/He ratio of 98/2 gives a better overall fit to both the \het\ and \lya\ measured absorption. 

From the analysis performed for the considered ranges of \mlr=10$^{8}$--10$^{12}$\,\gs\
 and temperature of 4000\,K to 11500\,K, we obtain the following results: 
(i) we find that the \het\ peak density is located in the altitude range of 1.04--1.60\,\rp; 
(ii) we obtain an [H]/[H$^{+}$] transition altitude ranging from about 1.2 to 1.9\,\rp; 
(iii) we obtain an effective radii of the XUV absorption in the range of 1.16--1.30\,\rp;
(iv) the averaged mean molecular weight of the gas ranges from 0.61 to 0.73\,g\,mole$^{-1}$;  
and (v) we find that the thermospheric H/He ratio should be larger than 90/10, the most likely value being about 98/2.

Thus, the \het\ absorption spectrum significantly constrains the thermospheric structure of \hd20 and advances our knowledge of its escaping atmosphere. In particular, we show that the H/He ratio should be larger than 90/10, that hydrogen is almost fully ionised ([H]/[H$^{+}$]\,<\,0.1) at altitudes above 2.9\,\rp, and that \het\ is accumulated at low thermospheric altitudes.  

Comparing our results with previous works we find that, overall, they generally agree well with the constrained \mlr(T) curve found in this work (orange curve in Fig.~\ref{chi2_2}). Only the values derived by \cite{Tian2005} 
and \cite{Koskinen2013a}, both larger than ours, fall relatively far from the derived constraints. 

Assuming the energy-limited approach, we derive the heating efficiency as a function of  the mass-loss rate (see Fig.~\ref{eta}). If we additionally assume the heating efficiency range of 0.1--0.2 derived from the detailed study of \cite{Shematovich_2014},  we conclude that the most probable substellar total mass-loss rate is in the range of (0.42--1.00)$\times 10^{11}$\,\gs, and that the temperature ranges from 7125 to 8125\,K.

Our model, as it is inherently limited by its one dimension and spherical symmetry, cannot explain any net (either blue or red) shifts. Future work on the analysis of the observed net blueshift components and on possible cometary outflows with 3D modelling is encouraged.

\begin{acknowledgements}
 We thank Prof. J. Linsky for his helpful refereed report.
 IAA authors acknowledge financial support from the State Agency for Research of the Spanish MCIU through the ``Center of Excellence Severo Ochoa" award SEV-2017-0709.
CARMENES is an instrument for the Centro Astronómico Hispano-Alemán de Calar Alto (CAHA, Almería, Spain). CARMENES is funded by the German Max-Planck-Gesellschaft (MPG), the Spanish Consejo Superior de Investigaciones Científicas (CSIC), the European Union through FEDER/ERF FICTS-2011-02 funds, and the members of the CARMENES Consortium (Max- Planck-Institut für Astronomie, Instituto de Astrofísica de Andalucía, Landessternwarte Königstuhl, Institut de Ciències de l'Espai, Insitut für Astrophysik Göttingen, Universidad Complutense de Madrid, Thüringer Landessternwarte Tautenburg, Instituto de Astrofísica de Canarias, Hamburger Sternwarte, Centro de Astrobiología and Centro Astronómico Hispano-Alemán), with additional contributions by the Spanish Ministry of Economy, the German Science Foundation through the Major Research Instrumentation Programme and DFG Research Unit FOR2544 “Blue Planets around Red Stars”, the Klaus Tschira Stiftung, the states of Baden-Württemberg and Niedersachsen, and by the Junta de Andalucía.
We acknowledge financial support from the Agencia Estatal de Investigaci\'on of the Ministerio de Ciencia, Innovaci\'on y Universidades, funds through projects:\,ESP2016--76076--R,\,ESP2017-87143-R,\,BES--2015--074542,\,BES--2015--073500,\,PGC2018-098153-B-C31,\,AYA2016-79425-C3-1/2/3-P.  
\end{acknowledgements}

\bibliographystyle{aa} 
\bibliography{ref.bib}

\begin{appendix}
\section{Calculation of the average mean molecular weight}\label{ap:mu}

From Eq.~(\ref{eq:momentum_cons_2}), considering the constant temperature $T_0$, we have
\begin{equation}
    \begin{split}
     \ve(r) \frac{d\ve}{dr}\,
     &+\,\frac{kT_{0}}{\mu(r)}\,\left( - \frac{1}{\ve(r)}\,\frac{d\ve}{dr}-\frac{2}{r}  \right) \\
     &+ k\,T_{0}\,\frac{d(1 /\mu)}{dr}
     + \frac{G\,M_{p}}{r^{2}} = 0 .
     \end{split}
    \label{eq:mmw_averaged_2}
\end{equation}

By integrating Eq.~(\ref{eq:mmw_averaged_2}) and re-arranging its terms, we obtain
\begin{equation}
    \begin{split}
    \displaystyle 
     \frac{G\,M_p}{kT_{0}}\, \int_{r_{0}}^{r_{f}} \mu(r)\, \frac{dr}{r^2}\, +
     & \frac{1}{kT_{0}}\, \int_{\ve_{0}}^{\ve_{f}} \mu(r)\,\ve(r)\, d\ve\,  +
     \int_{1/\mu_{0}}^{1/\mu_{f}}\mu(r)\,d(1 /\mu) \\
     &= \int_{\ve_{0}}^{\ve_{f}} \frac{d\ve}{\ve(r)} + 
     2 \int_{r_{0}}^{r_{f}}\frac{dr}{r},
     \end{split}
    \label{eq:mmw_averaged_1}
\end{equation}
where $r_{f}$ is the upper boundary distance of our model (10 \rp), and $\ve_{f}$ and $\mu_{f}$ are the bulk radial velocity and mean molecular weight, respectively.

The mean molecular weight, $\mu(r)$, enters in the first, second, and third terms of Eq.~(\ref{eq:mmw_averaged_1}). Thus, we construct an expression for $\overbar{\mu}$ : 
\begin{equation}
 \overbar{\mu} = \frac{\displaystyle 
     G\,M_p\, \int_{r_{0}}^{r_{f}} \mu(r)\, \frac{dr}{r^2}\, +
     \int_{\ve_{0}}^{\ve_{f}} \mu(r)\,\ve(r)\, d\ve\,  +
     kT_0\, \int_{1/\mu_{0}}^{1/\mu_{f}}\mu(r)\,d(1 /\mu)}
     {\displaystyle G\,M_p\, \int_{r_{0}}^{r_{f}} \frac{dr}{r^2}\, +
     \int_{\ve_{0}}^{\ve_{f}} \ve(r)\, d\ve\,  +
     kT_0\, \int_{1/\mu_{0}}^{1/\mu_{f}}\, d(1/\mu)}.
        \label{eq:mmw_averaged}
\end{equation}

\end{appendix}

\end{document}